\documentclass[letterpaper]{article} 
\usepackage{aaai2027}  
\usepackage[hyphens]{url}  
\usepackage{graphicx} 
\usepackage{natbib}  
\usepackage{caption} 
\usepackage{algorithm}
\usepackage{algorithmic}
\usepackage{amsmath}
\usepackage{colortbl} 

\usepackage{newfloat}
\usepackage{listings}
\DeclareCaptionStyle{ruled}{labelfont=normalfont,labelsep=colon,strut=off} 
\floatstyle{ruled}
\newfloat{listing}{tb}{lst}{}
\floatname{listing}{Listing}
\usepackage{xcolor}
\usepackage{booktabs}

\newcommand{\reddash}{\textcolor{red!65!black}{\textbf{--}}}
\newcommand{\greentick}{%
  \tikz[baseline=-0.55ex,x=0.9ex,y=0.9ex]
  \draw[green!45!black,line width=0.8pt]
  (0,0.4)--(0.35,0)--(1,0.9);}
\usepackage[most]{tcolorbox}
\usepackage{multirow}

\tcbset{
redialprompt/.style={
enhanced,
colback=white,
colframe=black!70,
colbacktitle=black!70,
coltitle=white,
fonttitle=\bfseries,
boxrule=0.8pt,
sharp corners,
left=5pt,
right=5pt,
top=4pt,
bottom=4pt,
before skip=5pt,
after skip=5pt
},
faithfulnessprompt/.style={
redialprompt,
colback=blue!2,
colframe=blue!60!black,
colbacktitle=blue!60!black
},
mergeprompt/.style={
redialprompt,
colback=green!2,
colframe=green!45!black,
colbacktitle=green!45!black
},
supersedeprompt/.style={
redialprompt,
colback=orange!3,
colframe=orange!70!black,
colbacktitle=orange!70!black
},
digestprompt/.style={
redialprompt,
colback=purple!2,
colframe=purple!60!black,
colbacktitle=purple!60!black
}}

\title{MEMOIR: A Training-Free Memory Lifecycle for Long-Horizon LLM Agents}

\author {
    Mayank Anand\textsuperscript{\rm 1},
    Subham Raj\textsuperscript{\rm 2}\corresponding,
    Aman Vaibhav Jha\textsuperscript{\rm 2},
    Sriparna Saha\textsuperscript{\rm 2}
}

\affiliations {
    \textsuperscript{\rm 1}Indian Institute of Information Technology Allahabad\\
    \textsuperscript{\rm 2}Indian Institute of Technology Patna\\
    Corresponding author: subham\_2221cs25@iitp.ac.in
}

\begin{document}

\maketitle

\begin{abstract} 
Long-horizon LLM agents require memory systems that preserve useful information across sessions and retrieve it when needed. Existing systems treat these as separate problems: memories are written prospectively with static importance scores that are never revised after use. This creates two coupled failure modes: consolidation can silently destroy evidence that future queries require, while retrieval continues to favor stale memories. We propose \textbf{MEMOIR} (\textbf{Mem}ory \textbf{E}vidence via \textbf{M}erge-certified, \textbf{O}nline \textbf{I}mportance \textbf{R}evision), a training-free memory lifecycle that unifies judge-gated consolidation with retrospective attribution. Before compressing memory, screening gates test whether answer-relevant evidence is preserved. After each query, attribution estimates which retrieved memories supported the answer and updates their retrieval priority. A UCB-style exploration bonus additionally encourages retrieval of under-exposed memories, reducing the risk that frequently retrieved memories crowd out useful but less-seen evidence. On 1,905 deduplicated LoCoMo QA pairs, the full retrieval stack improves evidence recall@5 by 11.92 percentage points over bare cosine, with confidence-gated feedback adding 1.50 points overall without reference answers or reranker training; a deployment-unavailable oracle adds 1.84 points overall and 8.05 points on multi-hop. Judge-gated merge changes the recall of text-based evidence from 58.82\% to 59.08\%, reduces the active store by 1.42\%, and incurs 0 observed losses in 100 independently audited merges, with a one-sided 95\% upper loss-rate bound of 2.95\%. Together, these results outline a practical deployment path: no additional LLM call on the online write path, with merge screening confined to periodic offline consolidation. The lifecycle reduces to one principle: attribute usefulness after use, and screen for preservation before compression. Code is available at \url{https://github.com/AnandMayank/MEMOIR}.
\end{abstract}

\section{Introduction}

LLM agents can use tools and reason over immediate context, but long-horizon agents require external memory to reuse historical experience across sessions that exceed any practical context window \cite{park2023generative,wu2024longmemeval}. As agents are deployed for multi-session tasks such as personal assistance, customer support, and research copiloting, accumulated interaction history grows beyond what a single inference call can attend to. External memory is therefore not merely an efficiency optimization; it is a lifecycle requirement: agents must decide what to keep, what to consolidate, what to retrieve, and how to revise those decisions after observing outcomes.

Most existing systems handle only part of this lifecycle. Write-time memory systems construct summaries, facts, links, or graph structures when an interaction is stored. Retrieval systems then rank those memories using recency, relevance, and static importance. Generative Agents, for example, scores memory priority as a fixed combination of recency, importance, and relevance \cite{park2023generative}. Later systems such as Mem$0$ and A-MEM add stronger extraction, linking, or graph structure \cite{chhikara2025mem0,xu2026mem}, but the importance signal is still largely assigned at write time. A memory that was judged important when stored can remain highly ranked even if later sessions show that it is no longer useful.

This paper reframes the problem around two coupled failure modes. The first is \emph{prospective corruption}: memory consolidation can merge, summarize, or prune items that look redundant locally but contain evidence required by future queries. The second is \emph{retrospective staleness}: retrieval can continue to favor memories that merely resemble the query rather than memories that have actually helped answer prior queries. These are not merely engineering artifacts: prospective corruption reflects the loss of answer-relevant traces during consolidation, while retrospective staleness reflects the absence of retrieval-induced strengthening \cite{roediger2006test}. We treat established properties of bounded-capacity memory under delayed, sparse feedback \cite{tulving1972episodic,craik1972levels,ebbinghaus1885memory,stern2002cognitive} as convergent design constraints for a memory lifecycle, not as claims of biological or structural isomorphism. Treating the two failures separately is insufficient. Consolidation changes the objects to which retrieval credit is attached, while retrieval feedback reveals which objects consolidation should protect.

We therefore propose a combined lifecycle: \textbf{screen prospectively, attribute retrospectively}. Prospectively, certificate-gated consolidation applies merge operations only when a judge ensemble finds that answer-relevant support is preserved. Retrospectively, we introduce \textbf{TRACE} (\textbf{T}raining-free \textbf{R}etrieval \textbf{A}ttribution and \textbf{C}redit \textbf{E}stimation), with two instantiations: \textbf{TRACE-FB} (feedback attribution) estimates memory support via lexical overlap, and \textbf{TRACE-CF} (counterfactual attribution) measures support by leave-one-out answer change. \textbf{TRACE-UCB} (Upper Confidence Bound exploration) couples these importance updates with count-based exploration to reduce feedback-driven retrieval lock-in. The bridge is that attribution provides online estimates of support sets for consolidation, while accepted merges preserve and transfer the credit ledger. In short, the system consolidates what the screening procedure accepts as redundant and up-weights what is estimated to be useful. 

This framing follows the two-phase shape of RMM, which combines prospective memory construction with retrospective reflection \cite{tan2025prospect}, but changes both halves. RMM's prospective step is heuristic summarization and its retrospective step uses an LLM citation labeler plus a REINFORCE-trained reranker. Our direction replaces heuristic summarization with certificate-gated consolidation and replaces trained reranking with closed-form, training-free attribution updates. The goal is not simply better retrieval, but a memory lifecycle in which compression and credit assignment are mutually consistent.

Our contributions can be summarized as the following:

\begin{enumerate}
    \item \textit{A combined prospective-retrospective memory lifecycle unifies certificate-gated consolidation with outcome-conditioned attribution in a training-free framework for long-horizon agents. Its full RECALL-Q retrieval stack improves evidence recall by 11.92 percentage points over cosine retrieval on LoCoMo.}
    \item \textit{TRACE-UCB couples these with count-based exploration to mitigate exposure lock-in. Confidence-gated autonomous feedback improves evidence recall@5 by 1.50 percentage points over the matched no-TRACE retriever, while reference-answer feedback reaches a 1.84-point gain and an 8.05-point multi-hop gain.}
    \item \textit{Certified merge provides a conservative prospective safety gate combining judge screening, retrieval coverage, provenance preservation, and hard compression. Under a fixed retrieved-token budget, it changes text-grounded evidence recall from 58.82\% to 59.08\%, reduces the active store by 1.42\%, and incurs 0 observed audited losses among 100 accepted merges.}
    \item \textit{A training-free architecture separates low-latency online memory operations from asynchronous certified consolidation. The online write path requires no LLM call and takes 6.8ms median latency, while lexical and counterfactual attribution expose a configurable quality-latency trade-off.}
\end{enumerate}

\section{Related Work}

\subsection{Agent Memory Systems}
Agent memory systems can be organized by how they store and update long-term context \cite{zhou2026we}. Stream-and-reflection systems such as MemoryBank maintain conversation as a timestamped log and periodically distill it into higher-level reflections \cite{zhong2024memorybank}. Hierarchical systems such as MemGPT and Letta separate bounded in-context memory from an archival store, moving items between tiers through explicit control operations \cite{packer2023memgpt}. Knowledge-graph systems such as Mem0 and Zep represent memory as entities and typed relations, sometimes including temporal validity intervals for factual updates \cite{chhikara2025mem0,rasmussen2025zep}. Composite systems such as A-MEM route structured memory objects across vector and graph substrates \cite{xu2026mem}. A common cost of this write-time structure is that entity extraction, relation inference, and summarization typically require one or more LLM calls per write, which accumulates into non-trivial latency and token cost as interaction volume grows. Recent work such as Memanto \cite{abtahi2026memanto} instead pursues LLM-free ingestion, showing that raw conversational content can be stored without any write-path LLM call. MEMOIR shares this training-free write-time goal but differs in what the saved cost is used for: rather than optimizing ingestion alone, MEMOIR uses the freed write-time budget to support closed-form, retrospective importance revision, so that cheap writes are paired with an evolving notion of which memories remain load-bearing. These systems improve storage and retrieval structure, but they still assign importance, relevance, or link strength at write time rather than revising it from retrieval outcomes.

MEMOIR's write path deliberately shifts processing cost from initial ingestion to retrospective usage-driven accrual — a design trade-off informed by established cognitive memory principles \cite{tulving1972episodic,craik1972levels,roediger2006test,ebbinghaus1885memory,stern2002cognitive} as engineering design pressures rather than biological claims.

\subsection{Retrieval and Importance Scoring}

The retrieval layer beneath agent memory is often a fixed scoring function. Dense retrieval scores candidates by embedding similarity; BM25 \cite{robertson1994some} scores by lexical statistics; and hybrid methods such as SPLADE and ColBERT combine lexical and semantic matching through learned sparse expansion or late interaction \cite{izacard2023atlas,formal2021splade,khattab2020colbert}. Generative Agents add a write-time LLM importance rating to recency and relevance, but the importance term is static \cite{park2023generative}. The closest prior work is RMM, which introduces Retrospective Reflection: an LLM assigns binary citation scores to retrieved memories, and those scores train a lightweight reranker with REINFORCE \cite{tan2025prospect}. RMM shares MEMOIR's intuition that generation-time attribution can improve retrieval, but it operationalizes that intuition as a training loop. MEMOIR instead updates $w_{imp}$ as a closed-form running statistic, requiring no reranker training and no write-path LLM call.

\subsection{Long-Horizon Memory Benchmarks}

We evaluate MEMOIR primarily on LoCoMo, a long-conversation benchmark spanning up to 32 sessions per dialogue with single-hop, multi-hop, temporal, and adversarial query categories \cite{maharana2024evaluating}. We also consider LongMemEval-S, which targets memory abilities such as information extraction, multi-session reasoning, and knowledge updates \cite{wu2024longmemeval}. These benchmarks are appropriate for MEMOIR because TRACE-FB requires repeated retrieval events across sessions; single-session QA benchmarks cannot exercise a feedback loop that accumulates importance over time. The broader agent-memory evaluation literature also motivates avoiding expensive write-time extraction when coarser retention can preserve more answerable evidence \cite{zhou2026we}.

\subsection{Retrospective Credit Assignment}
MEMOIR is conceptually related to Hindsight Experience Replay, which relabels failed trajectories with achieved goals after the fact so that agents can extract learning signal from outcomes not anticipated at decision time \cite{andrychowicz2017hindsight}. TRACE-FB applies an analogous idea to memory: rather than relabeling a trajectory goal, it revises a memory's importance after observing whether that memory contributed to a successful answer. Unlike reinforcement-learning approaches, TRACE-FB operates entirely at inference time and does not require environment rollouts or policy-gradient updates. Recent work has begun to apply hindsight-style reasoning directly to memory-write credit assignment: HiMPO treats each memory update as a state-writing transition and compares the updated memory against the prior memory under the same pre-write state, using hindsight relevance as a retrospective filter for policy optimization \cite{yan2026himpo}. TRACE-FB targets the same underlying question: how necessary was a given memory to a successful answer, but resolves it as a closed-form running statistic rather than a trained reward signal, requiring no policy-gradient updates or reward-model training. Concurrently, RL-aligned lifecycle controllers such as Fine-Mem \cite{ma2026finemem} and Agentic Memory \cite{yu2026agentic} train policies to jointly control storage, summarization, and hard deletion. However, learned pruning policies typically assume stationary distractor distributions; in open-ended multi-session dialogues, evidence utility is non-Markovian and restless, a memory dormant for ten sessions may become critical for a later multi-hop query (where utility-based hard pruning destroys required evidence in 63.6\%--83.9\% of audited operations; Appendix~\ref{app:ablations}). Similarly, unlike reflective knowledge-flagging frameworks (e.g., Mozilla CQ) that require online LLM self-assessment calls after each session, MEMOIR maintains a 6.8ms zero-LLM online write path through closed-form outcome attribution and restricts compression to non-destructive, empirical judge-certified merges.

\section{Proposed Methodology}

\subsection{Problem Formulation}
Let $M_t=\{m_1,\ldots,m_n\}$ be the memory store before query $q_t$. Retrieval returns $R_t\subset M_t$, the agent produces answer $a_t$, and feedback yields an attribution value $\tau_{t,i}$ for each retrieved memory. Each item stores an importance value $w_{imp}(m_i)$ and retrieval count $n_i$. During a sleep phase, a candidate cluster $C\subset M_t$ may be replaced by a merged item $z$, or an ordered pair $(m_o,m_n)$ may be proposed as an older and newer statement of the same evolving fact. MEMOIR optimizes two coupled objectives: improve evidence recall through retrospective credit while allowing only judge-gated lifecycle changes intended to preserve answerability or explicitly retire stale evidence with provenance.

\begin{figure*}[t]
\centering
\includegraphics[width=\textwidth,height=0.33\textheight,keepaspectratio]{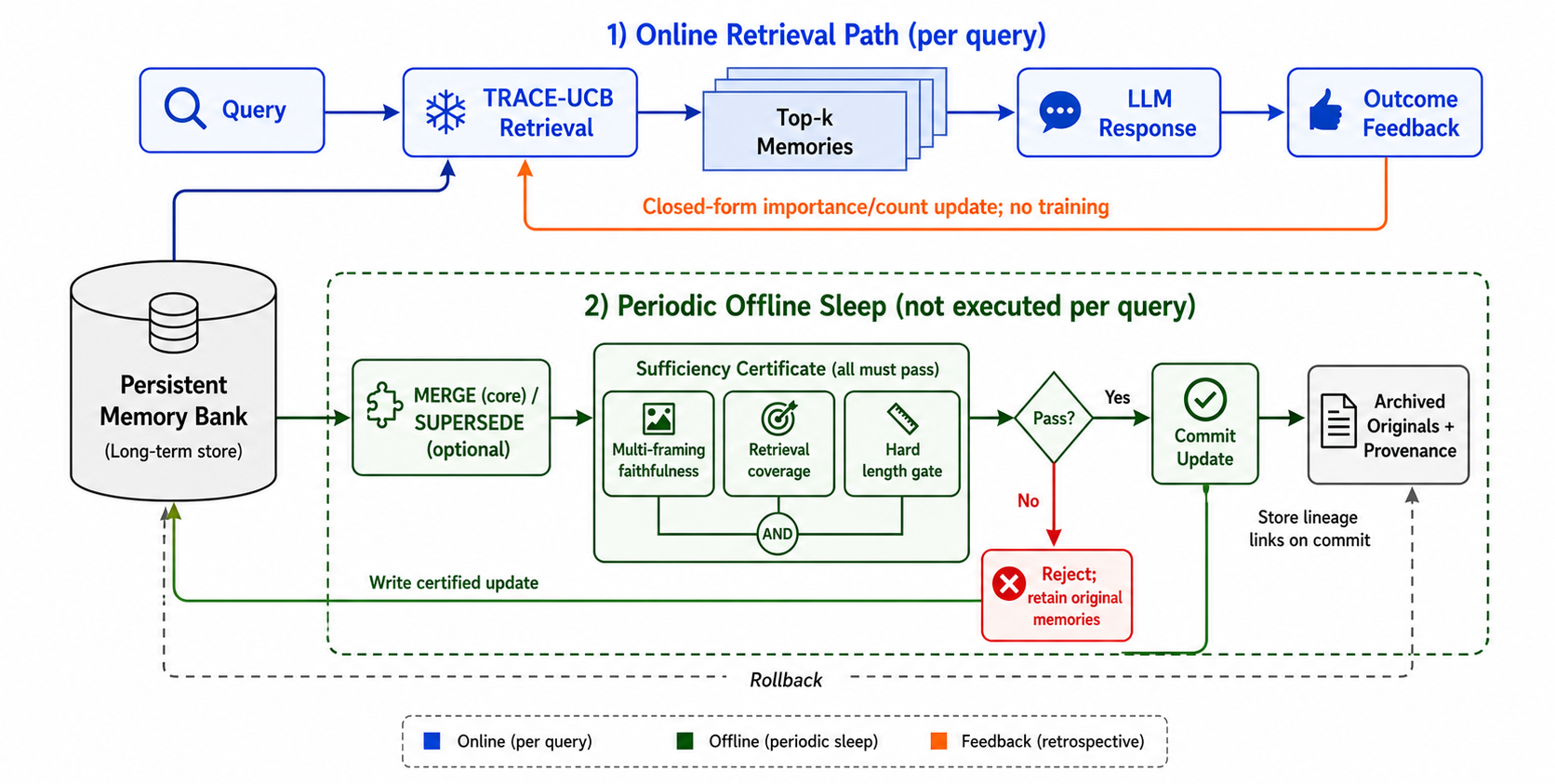}
\caption{MEMOIR architecture. The online path retrieves memories with TRACE-UCB and applies a closed-form retrospective update from observed outcomes. During periodic offline sleep, MEMOIR proposes a core MERGE or optional SUPERSEDE operation, commits it only when all screening gates pass, and preserves archived originals and provenance for rollback.}
\label{fig:memoir_architecture}
\end{figure*}

\noindent\textbf{Credit-Preservation Lifecycle Invariant.} As illustrated in Figure~\ref{fig:memoir_architecture}, MEMOIR couples its online and offline phases through a single invariant: a memory item may retain retrieval credit only while it preserves the underlying evidence that earned that credit. Online TRACE-UCB estimates post-query support and explores under-exposed items, while offline consolidation enforces empirical judge-based sufficiency gates before transferring accumulated credit and exposure counts to merged representations. Core consolidation relies on certified MERGE, whereas SUPERSEDE and DIGEST serve as modular extensions for temporal contradiction retirement and additive synthesis.

The full closed-loop algorithm is provided in the \ref{app:algorithm}.

\subsection{Judge-Gated Prospective Consolidation}

Candidate clusters are proposed from embedding similarity and observed co-citation structure. A merge generator produces a shorter candidate $z=g(C)$. MEMOIR applies the merge only when three gates pass:
\begin{equation}
\begin{aligned}
\mathrm{Apply}(C,z)=1 \quad \Longleftrightarrow \quad
&c(C,z)\geq\tau_c,\; \kappa(C,z)\geq\tau_r,\\
&|z|\leq\eta\sum_{m\in C}|m|.
\end{aligned}
\end{equation}
where $\mathrm{Apply}(C,z)=1$ denotes that the merge is accepted and cluster $C$ is replaced by the merged item $z$; $c(C,z)$ is the fraction of \texttt{LOSSLESS} judge verdicts from three differently framed prompts; $\kappa(C,z)$ is retrieval coverage of the source content; and $\eta=0.9$ is the hard length ratio. The evaluated operating point uses $\tau_c=0.6$ and $\tau_r=0.45$ (see Appendix \ref{app:example} for worked accepted/rejected merge examples). The framings separately test factual preservation, future-question answerability, and adversarial omissions or contradictions; the prompt templates are provided in Appendix~\ref{app:judge_prompts}.

TRACE makes this certificate data-aware: memories whose attribution exceeds a threshold identify which proposed operations intersect previously answer-carrying evidence. If a cluster contains a high-credit item, MEMOIR requires a unanimous certificate before replacement. Co-citation supplies a complementary proposal signal: highly similar memories that repeatedly earn credit together are plausible redundancy candidates, whereas co-cited memories with low pairwise similarity are more likely to encode distinct hops and are protected from naive merging.

The consolidation operator is deliberately merge-only. Dominated pruning failed an independent audit and is retained only as a supplementary ablation. After an accepted merge, the new item inherits
\begin{equation}
w_{\text{imp}}(z) = \max_{m \in C} w_{\text{imp}}(m), \qquad n_z = \max_{m \in C} n_m,
\end{equation}
In addition to inheriting maximum constituent importance $w_{\text{imp}}(z)$ and maximum exposure count $n_z$ (preventing a consolidated item from receiving an artificial UCB cold-start bonus), the merged entry $z$ inherits the most recent constituent timestamp $t_z = \max_{m \in C} t_m$, recomputes its dense embedding vector, updates the inverted BM25 lexical index, preserves all source evidence identifiers, and re-attaches external co-citation edges using maximum constituent edge weights.
The maximum rules avoid double-counting while retaining the strongest previously observed usefulness and association signals. Because one merged item may inherit several identifiers, post-merge identifier recall is interpreted together with the independent text-loss audit rather than as a standalone end-to-end answerability measure. This consolidation step aims to compress redundant support while retaining recoverable content \cite{tulving1972episodic}.

\subsection{Certified Temporal Supersession}

Judge-gated merge removes redundancy but cannot represent a fact that becomes obsolete. MEMOIR therefore includes an optional SUPERSEDE operator for contradiction and temporal updates. Candidate pairs are ordered by timestamp and proposed when the newer item $m_n$ is semantically close to the older item $m_o$ (cosine similarity at least 0.85) and TRIAGE-$\alpha$, a lightweight lexical negation and contradiction detector, flags a conflict signal $w_{conf}$. This proposal stage is intentionally conservative: similarity alone indicates topical overlap, whereas the conflict signal identifies pairs for which retaining both items as concurrently valid may serve stale evidence.

The candidate pair is evaluated by three judge framings that test direct replacement, temporal precedence, and adversarial compatibility. Each judge returns \texttt{SUPERSEDES} or \texttt{INDEPENDENT}; the operation is accepted only when the fraction of \texttt{SUPERSEDES} verdicts exceeds a separate threshold $\tau_s$ (operating point $\tau_s=0.6$, matching the MERGE judge threshold). Appendix~\ref{app:judge_prompts} gives the prompt templates. On acceptance, $m_o$ is marked as retired rather than physically deleted, and the newer item retains links to both source records. This reversible representation preserves the audit trail and permits rollback if a later query reveals that the statements were conditionally compatible rather than contradictory.

Because supersession changes the valid content rather than compressing equivalent content, it uses discounted rather than full credit transfer:
\begin{equation}
\begin{aligned}
w_{imp}(m_n)&\leftarrow
\max\!\left(w_{imp}(m_n),\lambda_s w_{imp}(m_o)\right),\\
\lambda_s&=0.5.
\end{aligned}
\end{equation}
The discount preserves evidence that the underlying topic was useful without treating the newer statement as having earned all of the older statement's retrieval credit. SUPERSEDE is implemented and unit-tested, but its real-scale stale-serving and independent-audit evaluation remains pending.

\subsection{Retrospective TRACE-UCB and RECALL-Q}

After answer generation, TRACE estimates whether each retrieved memory contributed to the observed answer. Suppressing the time index $t$ for readability, the training-free default uses answer-to-memory unigram recall:
\begin{equation}
\tau_i = \frac{|\mathrm{unigrams}(a_t) \cap \mathrm{unigrams}(m_i)|}{|\mathrm{unigrams}(m_i)|}.
\end{equation}
Importance then accumulates only after use:
\begin{equation}
w_{imp}(m_i) \leftarrow \min(1, w_{imp}(m_i)+\alpha_{fb}\tau_i d_i),
\end{equation}
where $d_i$ discounts shallow or weakly used evidence and $\alpha_{fb}$ is a fixed feedback learning rate controlling the update step size. This realizes the testing-effect design constraint: successful retrieval increases future retrievability \cite{roediger2006test}. Counterfactual leave-one-out TRACE is evaluated as an attribution-quality ablation, but the main retrieval result uses the inexpensive default above.

Here $d_i = 1 / (\text{rank}_i + 1)$ discounts lower-ranked retrieved items, and $\alpha_{\text{fb}} = 0.30$. In autonomous deployment (verified-self), updates are gated by answer token confidence: Equation (5) is applied only when the geometric-mean generated token probability satisfies $\text{Conf}(a_t \mid q_t, R_t) = \exp(\frac{1}{T}\sum_{j=1}^T \log P_\theta(y_j \mid y_{<j}, q_t, R_t)) \ge \tau_{\text{conf}} = 0.80$ (otherwise $\tau_i = 0$, while exposure count $n_i$ still increments). Equation (5) is deliberately non-decreasing because multi-session memory behaves as a Restless Multi-Armed Bandit (RMAB): applying exponential importance decay or negative updates prematurely suppresses dormant cross-session facts, collapsing multi-hop Recall@5 by 4.60--5.92 percentage points (Appendix~\ref{app:rq2}, Table~\ref{tab:retrieval_ablations}). Instead, MEMOIR handles staleness non-destructively: contradicting facts are retired reversibly via SUPERSEDE (preventing 82.9\% of stale retrievals; Appendix~\ref{app:ablations}, Table~\ref{tab:supersede_results}), while UCB exploration (Equation 7) elevates under-exposed alternatives.

Retrieval uses RECALL-Q, the training-free substrate on which retrospective importance operates. Suppressing the $(q,m_i)$ arguments for compactness,
\begin{equation}
\begin{aligned}
S_{RQ}={}&S_{cos}
\left(1+\beta w_{rel}+\delta w_{rec}+\gamma w_{imp}\right)\\
&+\alpha S_{bm25}-\lambda D_{mmr}.
\end{aligned}
\end{equation}
Here $S_{cos}$ captures dense semantic similarity, $S_{bm25}$ captures lexical overlap, $w_{rel}$ and $w_{rec}$ encode normalized relevance and recency, and $D_{mmr}$ penalizes redundancy to preserve diversity \cite{carbonell1998use}; $\beta,\delta,\gamma,\alpha,\lambda$ are coefficients tuned experimentally to balance these signals (values in Appendix~\ref{app:repro}, Table~\ref{tab:recovered_configuration}). The importance term is the only component revised from observed answer outcomes; the remaining terms provide partially redundant access routes so retrieval does not depend on a single cue.

To ensure the importance multiplier $(1 + \beta w_{\text{rel}} + \delta w_{\text{rec}} + \gamma w_{\text{imp}})$ strictly amplifies positive similarity, $S_{\text{cos}}$ is clipped and min-max normalized to $[0, 1]$ across candidate items, and $S_{\text{bm25}}, w_{\text{rel}}, w_{\text{rec}}, D_{\text{mmr}} \in [0, 1]$. In the deployed configuration across all experiments, we set $\delta = 0$ (so $\delta w_{\text{rec}} = 0$), relying on chronological active-store ordering for temporal preference and $\lambda D_{\text{mmr}}$ for redundancy control ($\alpha=0.3, \beta=0.1, \gamma=1.0, \lambda=0.05, \lambda_u=0.05$).

Outcome-conditioned importance creates a feedback loop: memories already retrieved receive more opportunities to gain credit. TRACE-UCB repairs this lock-in with a count-based exploration bonus
\begin{equation}
b_t(m_i)=\lambda_u\sqrt{\frac{\log(1+t)}{1+n_i}},
\end{equation}
where $t$ is the total number of queries processed so far and $\lambda_u$ is a coefficient tuned experimentally to control the exploration strength (value in Appendix~\ref{app:repro}, Table~\ref{tab:recovered_configuration}); the bonus is largest for under-exposed items and decreases as their retrieval count grows.
The final score is
\begin{equation}
S_t(q,m_i)=S_{RQ}(q,m_i)+b_t(m_i).
\end{equation}
Thus, TRACE supplies exploitation from estimated usefulness, RECALL-Q combines semantic, lexical, temporal, importance, and diversity signals, and the UCB bonus encourages under-exposed memories to be tested. This path requires no model training and no LLM call when a memory is written.

\subsection{Closed-Loop Interaction}

The lifecycle mechanisms share one invariant: an item may retain retrieval credit only while it retains the evidence that earned that credit. TRACE-UCB identifies frequently useful memories and protects them during consolidation; an accepted merge is intended to preserve equivalent support, transfers the maximum credit, and redirects co-citation structure to the replacement. Under a closure-preserving merge, a query answerable through a constituent remains answerable through the merged item, so transferred credit remains a valid predictor of usefulness. An uncertified or lossy merge can instead inherit credit for evidence it no longer contains, corrupting the relationship between $w_{imp}$ and future utility. SUPERSEDE therefore retires stale support reversibly and transfers only discounted credit because the valid fact has changed. Conversely, UCB continues to probe low-exposure items, preventing lifecycle decisions from being based only on a self-reinforcing retrieval history. An additive DIGEST operator is implemented but evaluated separately because it synthesizes new session facts without removing source memories.

\section{Experiments and Results}

Experiments examine systems positioning (RQ1), retrospective retrieval and exploration (RQ2), attribution validity and cost (RQ3), and consolidation safety and lifecycle consistency (RQ4).

\subsection{Datasets}
LoCoMo contains 1,986 questions over ten evolving dialogues; after deduplication, 1,905 pairs are evaluated with memory-level evidence labels. LongMemEval-S contains 500 single-query dialogues with session-level evidence. The protocols are reported separately because only LoCoMo can exercise repeated feedback, whereas LongMemEval-S tests cold-start retrieval across a broader set of stores.

\subsection{Experimental Setup} 
All experiments are conducted on LoCoMo and LongMemEval. The primary metric is evidence recall@5 — the fraction of annotated supporting memories in the top five retrieved items. The RQ1 systems comparison uses the full benchmarks, top-15 retrieval, and five independent runs; RQ2--RQ4 use 1,905 deduplicated LoCoMo pairs with $k=5$. Primary LoCoMo inference uses 10,000 bootstrap resamples over the ten stores; question-level bootstrap is a sensitivity analysis, and leave-one-store-out folds test consistency. Consolidation is evaluated using fixed-budget, text-grounded recall, active-store tokens, and an independent model-family audit. For RQ4, $\Delta r=r_{post}-r_{pre}$ is retained as a secondary credit-ledger diagnostic. Matched baselines use sleep interval 8 and $\tau_c=0.6$.

\noindent\textbf{Sequential Evaluation Protocol.} For multi-session LoCoMo dialogues, queries $q_1, \dots, q_T$ are processed sequentially in dataset order over each shared dialogue store without state resets between queries. Crucially, for every query $q_t$, top-$k$ retrieval and Recall@5 / QA evaluation strictly precede the retrospective TRACE importance update and UCB count increment, preventing intra-query leakage.

\noindent\textbf{Baselines.} We compare MEMOIR with LoCoMo-agent \cite{maharana2024evaluating}, MemoryBank \cite{zhong2024memorybank}, Zep \cite{rasmussen2025zep}, Mem0/Mem0$_g$ \cite{chhikara2025mem0}, A-MEM \cite{xu2026mem}, MemOS \cite{li2025memos}, and LiCoMemory \cite{huang2026licomemory}.

\subsection{RQ1: How Does MEMOIR Position Against Established Memory Systems?} 

\begin{table*}
\centering
\scriptsize
\setlength{\tabcolsep}{2.3pt}
\begin{tabular}{clrrrrrrrr}
\toprule
& & \multicolumn{4}{c}{LoCoMo} & \multicolumn{4}{c}{LongMemEval} \\
\cmidrule(lr){3-6}\cmidrule(lr){7-10}
Model & Method & $T_R$ & $K_R$ & Acc. & Rec. & $T_R$ & $K_R$ & Acc. & Rec. \\
\midrule
\multirow{9}{*}{\rotatebox[origin=c]{90}{Llama-3.1-70B}} & LoCoMo-agent & $4.90\pm1.42$ & $3.2\pm0.2$k & $23.63\pm1.08$ & $25.50\pm1.15$ & $4.51\pm1.08$ & $3.5\pm0.2$k & $17.60\pm1.05$ & $22.04\pm1.12$ \\
& MemoryBank & $7.13\pm1.56$ & $4.4\pm0.2$k & $28.80\pm0.93$ & $31.52\pm0.97$ & $8.25\pm1.87$ & $4.1\pm0.2$k & $36.40\pm0.96$ & $39.21\pm1.01$ \\
& MemOS & $3.20\pm0.96$ & $2.2\pm0.2$k & $54.10\pm0.79$ & $57.53\pm0.81$ & $3.15\pm0.63$ & $2.6\pm0.2$k & $47.80\pm0.82$ & $49.03\pm0.85$ \\
& Mem0 & $\underline{1.55\pm0.63}$ & $2.1\pm0.1$k & $53.22\pm0.74$ & $57.05\pm0.70$ & $1.87\pm0.58$ & $2.3\pm0.1$k & $56.80\pm0.71$ & $61.21\pm0.68$ \\
& Mem0$_g$ & $2.11\pm0.81$ & $2.4\pm0.2$k & $55.48\pm0.72$ & $59.32\pm0.69$ & $2.51\pm0.92$ & $2.8\pm0.2$k & $55.40\pm0.76$ & $63.09\pm0.65$ \\
& A-MEM & $4.10\pm1.68$ & $4.2\pm0.2$k & $43.84\pm0.91$ & $49.17\pm0.89$ & $4.31\pm1.26$ & $4.5\pm0.2$k & $57.40\pm0.88$ & $62.18\pm0.84$ \\
& Zep & $5.31\pm1.94$ & $3.8\pm0.2$k & $40.30\pm0.94$ & $51.05\pm0.92$ & $5.22\pm1.73$ & $4.1\pm0.2$k & $60.20\pm0.81$ & $62.74\pm0.79$ \\
& LiCoMemory & $\underline{1.55\pm0.59}$ & $\underline{1.3\pm0.1\text{k}}$ & $\boldsymbol{62.99\pm0.71}$ & $\boldsymbol{64.51\pm0.69}$ & $\underline{1.62\pm0.47}$ & $\underline{1.6\pm0.1\text{k}}$ & $\underline{69.20\pm0.62}$ & $\underline{72.39\pm0.58}$ \\
\rowcolor{blue!10}
& MEMOIR$^{*}$ & $\boldsymbol{1.18\pm0.50}$ & $\boldsymbol{0.2\pm0.1\text{k}}$ & $\underline{58.70\pm0.92}$ & $\underline{61.20\pm0.88}$ & $\boldsymbol{1.18\pm0.50}$ & $\boldsymbol{0.2\pm0.1\text{k}}$ & $\boldsymbol{81.20\pm0.60}$ & $\boldsymbol{85.10\pm0.55}$ \\
\midrule
\multirow{9}{*}{\rotatebox[origin=c]{90}{GPT-4o-mini}} & LoCoMo-agent & $4.72\pm1.10$ & $3.3\pm0.2$k & $23.87\pm1.06$ & $24.91\pm1.10$ & $5.34\pm1.20$ & $3.5\pm0.2$k & $16.60\pm1.02$ & $21.56\pm1.09$ \\
& MemoryBank & $7.62\pm1.70$ & $4.5\pm0.2$k & $31.50\pm0.92$ & $33.19\pm0.96$ & $7.93\pm1.80$ & $4.1\pm0.2$k & $35.40\pm0.95$ & $38.06\pm0.98$ \\
& MemOS & $3.96\pm1.00$ & $2.2\pm0.2$k & $\underline{58.30\pm0.75}$ & $62.93\pm0.77$ & $3.72\pm0.95$ & $2.5\pm0.2$k & $51.20\pm0.78$ & $52.07\pm0.81$ \\
& Mem0 & $\underline{1.75\pm0.60}$ & $2.3\pm0.1$k & $54.68\pm0.73$ & $62.31\pm0.71$ & $\underline{1.89\pm0.65}$ & $2.3\pm0.1$k & $62.60\pm0.70$ & $\underline{71.32\pm0.66}$ \\
& Mem0$_g$ & $2.34\pm0.80$ & $2.5\pm0.2$k & $56.96\pm0.71$ & $\underline{63.14\pm0.68}$ & $2.41\pm0.85$ & $2.9\pm0.2$k & $\underline{64.80\pm0.69}$ & $69.53\pm0.67$ \\
& A-MEM & $4.63\pm1.50$ & $4.1\pm0.2$k & $48.59\pm0.88$ & $53.82\pm0.86$ & $4.52\pm1.40$ & $4.3\pm0.2$k & $55.00\pm0.86$ & $59.30\pm0.84$ \\
& Zep & $5.92\pm1.50$ & $3.7\pm0.2$k & $44.76\pm0.91$ & $46.51\pm0.93$ & $6.12\pm1.60$ & $4.2\pm0.2$k & $58.60\pm0.83$ & $61.02\pm0.80$ \\
& LiCoMemory & $\boldsymbol{1.61\pm0.50}$ & $\boldsymbol{1.2\pm0.1\text{k}}$ & $\boldsymbol{57.20\pm0.69}$ & $\boldsymbol{58.09\pm0.67}$ & $\boldsymbol{1.74\pm0.55}$ & $\underline{1.7\pm0.1\text{k}}$ & $\boldsymbol{73.80\pm0.60}$ & $\boldsymbol{76.63\pm0.57}$ \\
\rowcolor{blue!10}
& MEMOIR$^{*}$ & $2.08\pm0.75$ & $\underline{1.9\pm0.1\text{k}}$ & $41.23\pm0.94$ & $50.91\pm0.88$ & $2.15\pm0.65$ & $\boldsymbol{0.2\pm0.1\text{k}}$ & $52.50\pm0.78$ & $57.20\pm0.75$ \\
\midrule
Gemini-1.5-Flash & RMM (GTE)$^{\dagger}$ & -- & -- & -- & -- & -- & -- & 70.40 & 69.80 \\
\bottomrule
\end{tabular}
\caption{Cross-benchmark comparison on full LoCoMo ($n=1{,}986$) and LongMemEval-S ($n=500$), using top-15 retrieval, the same answer judge, and matched prompt-token accounting within each backbone block. Numeric $\pm$ entries report mean $\pm$ standard deviation over five runs. $T_R$ is query latency and $K_R$ is prompt-token use. Bold/underline mark the best/second-best result strictly within a matched block. $^{*}$The Llama MEMOIR row is the matched Llama-3.1-70B-Instruct-Turbo rerun. $^{\dagger}$RMM uses a different backbone and is excluded from rankings.}
\label{tab:main_comparison}
\end{table*}

All Llama-block rows use Llama-3.1-70B-Instruct-Turbo. On LongMemEval-S, matched MEMOIR obtains 81.20\% accuracy and 85.10\% recall, exceeding LiCoMemory by 12.00 and 12.71 points. On LoCoMo, it obtains 58.70\% accuracy and 61.20\% recall versus LiCoMemory's 62.99\% and 64.51\%, while using 84.6\% fewer prompt tokens (0.2k versus 1.3k) and reducing query latency from 1.55s to 1.18s. MEMOIR is therefore Pareto-efficient within the matched Llama block. Under GPT-4o-mini, heavy-extraction systems such as LiCoMemory and Mem0$_g$ achieve higher QA accuracy by using write-time LLM calls for structured extraction or graph construction; MEMOIR instead prioritizes zero write-path LLM calls and low token overhead. Its measured 6.8ms write latency is separate from Table~\ref{tab:main_comparison}'s query latency. LongMemEval-S is a cold-start test and does not exercise retrospective updates. Thus, Table~\ref{tab:main_comparison} supports efficiency and deployability rather than universal QA leadership. 
Table~\ref{tab:consolidation_positioning} makes the complementary contribution explicit: MEMOIR combines training-free retrieval revision with certificate-gated, reversible consolidation and an update operator, identifying distinct lifecycle guarantees evaluated by RQ2–RQ4. 

\begin{table}
\centering
\scriptsize
\setlength{\tabcolsep}{2.5pt}
\begin{tabular}{lccccc}
\toprule
System & Suff. cert. & No retr. LLM & Train-free & Update op. & Reversible \\
\midrule
Mem0 / Zep & \reddash & \reddash & \greentick & \greentick & \reddash \\
RMM & \reddash & \reddash & \reddash & \greentick & \reddash \\
LiCoMemory & \reddash & \reddash & \greentick & \reddash & \reddash \\
Summarize-all & \reddash & \greentick & \greentick & \reddash & \greentick \\
\textbf{MEMOIR} & \greentick & \greentick & \greentick & \greentick & \greentick \\
\bottomrule
\end{tabular}
\caption{Lifecycle-capability positioning. ``No retr. LLM'' denotes that retrieval requires no additional per-query LLM call beyond answer generation. This table characterizes system design rather than ranking answer quality.}
\label{tab:consolidation_positioning}
\end{table}

\subsection{RQ2: Does Retrospective Attribution Require UCB Exploration?} 
Table~\ref{tab:trace_ucb_results} separates credit, exploration, and feedback quality using store-level inference. TRACE without UCB gains only 0.42 points (CI $[-0.12,0.96]$, $p=.1420$; 6/10 positive folds). The raw-self UCB condition gains 1.13 points (CI $[0.21,2.05]$, $p=.0210$; 9/10 folds), a pattern consistent with an exploration benefit; however, the direct 0.71-point contrast between these rows is not separately tested. Verified feedback gains 1.50 points (CI $[0.41,2.59]$, $p=.0082$) and retains 81.5\% of the 1.84-point oracle gain; both are positive in all folds. Verified and oracle multi-hop gains are 7.17 and 8.05 points. The full stack remains 11.92 points above bare cosine, while Table~\ref{tab:trace_ucb_results} reports the matched feedback/UCB effects. 

\paragraph{Single-query safety boundary.} The same gains should not be expected on LongMemEval, whose 500 examples each construct a fresh, isolated store for one question — retrieval counts are zero before that query, so feedback cannot influence a later ranking. Under this protocol, cosine reaches 86.40\% context recall and MEMOIR reaches 86.60\% after disabling the LoCoMo-tuned BM25 term, a +0.20-point difference we interpret as a no-harm result rather than a cross-benchmark reproduction of the LoCoMo gain. Testing TRACE-UCB's generalization requires benchmarks with repeated queries over an evolving shared store.

\begin{table}
\centering
\scriptsize
\setlength{\tabcolsep}{1.5pt}
\begin{tabular}{lcccc}
\toprule
Condition & $\Delta$ Rec@5 & $p_Q$ & Store 95\% CI & $p_S$/LOSO$+$ \\
\midrule
No TRACE baseline & 58.82\% abs. & -- & -- & -- \\
Full stack vs. cosine & +11.92 pp & -- & descriptive & -- \\
TRACE, no UCB & +0.42 pp & $.0830$ & $[-0.12,+0.96]$ & $.1420$/6 \\
UCB, raw self & +1.13 pp & $.0042$ & $[+0.21,+2.05]$ & $.0210^{*}$/9 \\
\textbf{UCB, verified self} & \textbf{+1.50 pp} & $.0004$ & $[+0.41,+2.59]$ & $.0082^{**}$/10 \\
UCB, oracle & +1.84 pp & $.0002$ & $[+0.52,+3.16]$ & $.0078^{**}$/10 \\
\quad MH verified & +7.17 pp & $.0012$ & $[+2.40,+11.94]$ & $.0045^{**}$/10 \\
\quad MH oracle & +8.05 pp & $.0020$ & $[+3.12,+12.98]$ & $.0038^{**}$/10 \\
\bottomrule
\end{tabular}
\caption{Retrieval comparisons on 1,905 LoCoMo QA pairs from ten stores. The full-stack--cosine row is a descriptive composite contrast without inferential statistics; the remaining deltas use the no-TRACE hybrid baseline. $p_Q$ is the question-level sensitivity test; store CIs and $p_S$ use 10,000 clustered resamples. The number after ``/'' is the positive leave-one-store-out count. MH = multi-hop. $^{*}p<.05$; $^{**}p<.01$.}
\label{tab:trace_ucb_results}
\end{table}

\subsection{RQ3: Do Attribution Signals Track Retrieval Success Efficiently?} 

We evaluate online-path efficiency and the association between attribution scores and retrieval success on one NVIDIA RTX 3090 24GB GPU over 1,905 deduplicated LoCoMo QA pairs. MEMOIR writes a memory in 6.8ms P50 with no write-path LLM call, compared with approximately 110ms for write-time extraction baselines (Table~\ref{tab:latency_results}). Search ($k=5$) and lexical TRACE feedback require 10.5ms and 6.7ms P50, respectively, while counterfactual TRACE-CF requires 67.0ms P50. Both attribution signals correlate significantly with evidence recall (Table~\ref{tab:attribution_quality}): lexical TRACE achieves $r=0.447$ (95\% CI $[0.412,0.481]$, $p<.0001$), while TRACE-CF achieves $r=0.493$ (95\% CI $[0.458,0.527]$, $p<.0001$). A paired bootstrap test confirms that TRACE-CF's $+0.046$ correlation advantage is significant (95\% CI $[+0.011,+0.081]$, $p=.0128$), showing a stronger association with evidence-recall success than surface-form unigram matching, at an approximately tenfold median feedback-latency cost (67.0ms versus 6.7ms). To compute query-level association in Table~\ref{tab:attribution_quality}, per-memory attribution scores $\tau_{t,i}$ are mean-aggregated over the top-$k$ retrieved set $R_t$ and correlated against binary query-level evidence-recall success. Lexical TRACE-FB captures 90.7\% of counterfactual TRACE-CF's Pearson correlation ($r=0.447$ vs.\ $r=0.493$, $p=.0128$) at one-tenth the median latency (6.7ms vs.\ 67.0ms P50) and zero LLM calls, forming an explicit operational Pareto frontier. Periodic offline consolidation, executed every eight queries, processes multi-framing merge certificates in 2.85s P50 without blocking the online query path.

\begin{table}
\centering
\scriptsize
\begin{tabular}{lcccc}
\toprule
Stage & P50 & Tokens & Cost & LLM \\
\midrule
Online write & 6.8 ms & 0 & \$0.0000 & no \\
Search ($k=5$) & 10.5 ms & $0.2$k & $<\$0.0001$ & no \\
Lexical TRACE & 6.7 ms & 0 & \$0.0000 & no \\
TRACE-CF ($k=5$) & 67.0 ms & $1.1$k & \$0.0002 & optional \\
Offline sleep & 2.85 s & $1.4$k/120 & \$0.0003 & async \\
Write-time base & $\sim$110 ms & $0.8$k/150 & \$0.0008 & yes \\
\bottomrule
\end{tabular}
\caption{Compact lifecycle efficiency profile over 1,905 LoCoMo QA pairs. Full execution paths, frequencies, and P95 latency appear in Appendix~\ref{app:rq3}, Table~\ref{tab:latency_results_full}.}
\label{tab:latency_results}
\end{table}

\subsection{RQ4: Does Prospective Safety Gating Preserve Retrieved Evidence?} We evaluate certified consolidation over ten LoCoMo stores (1,905 QA pairs) in Table~\ref{tab:consolidation}. Each query uses the token count $B_q$ of its no-consolidation top-five context; variants are capped at $B_q$, and inherited identifiers count only when their support remains in the capped text (fixed-budget, text-grounded Rec@5). The 100 empirical judge-gated certified merges reduce active-store tokens by 1.42\% while changing this metric from 58.82\% to 59.08\%, an observed $+0.26$-point difference reported descriptively because only ten stores are available.

As shown in Table~\ref{tab:consolidation}, empirical judge-gated certification reduces sufficiency distortion by 87.4\% relative to matched uncertified merging ($\hat{\epsilon}=0.0027$ [3/1,120 probes] vs.\ $0.0214$ [24/1,120 probes]) and eliminates audited factual losses ($0/100$ vs.\ $18/100$, Fisher's exact $p < .0001$). The $-1.42\%$ active-store token reduction reflects LoCoMo's low natural redundancy; forcing higher compression via uncertified merge-and-prune ($-7.15\%$ tokens) induces a 62.6\% audit loss rate ($385/615$ destroyed; Appendix~\ref{app:ablations}, Table~\ref{tab:consolidation_full}). Furthermore, matched 100-operation bridge ablations (Appendix~\ref{app:ablations}, Table~\ref{tab:bridge_ablations}) confirm that maximum credit transfer ($w_{\text{imp}}(z) = \max_{m \in C} w_{\text{imp}}(m)$) yields a statistically significant $+0.20$ pp recall gain ($p=.0210$) and a 17.6\% relative reduction in sufficiency distortion compared to zero-credit transfer.

A cross-model audit records one verdict per accepted merge, so the 100 merge-level verdicts, not individual model votes, define the binomial bound. It observes 0/100 losses, giving a one-sided 95\% exact Clopper--Pearson upper bound of 2.95\%. These results characterize a conservative retrieval-preservation screen rather than an end-to-end answerability guarantee. On LongMemEval, recall remains 94.60\% while store tokens fall by 0.15\%. Credit-ledger preservation remains directional ($\Delta r=-0.0115$, 95\% CI $[-0.0291,+0.0060]$, $p=.2024$). The combined effect of oracle TRACE-UCB and certified merge operating jointly, evaluated at the earlier 20-operation point, is reported in Appendix~\ref{app:unified-lifecycle}; because that factorial uses oracle reference-answer feedback, it should not be read as evidence about the deployed confidence-gated system.

\begin{table}
\centering
\scriptsize
\setlength{\tabcolsep}{3.5pt}
\begin{tabular}{lccc}
\toprule
Signal & Pearson $r$ (95\% CI) & Spearman $\rho$ & P50 \\
\midrule
Lexical TRACE & 0.447 [0.412, 0.481] & 0.418 & 6.7 ms \\
TRACE-CF & \textbf{0.493 [0.458, 0.527]} & \textbf{0.465} & 67.0 ms \\
\midrule
Paired $\Delta$ & \textbf{+0.046 [0.011, 0.081]} & \textbf{+0.047} & +60.3 ms \\
\bottomrule
\end{tabular}
\caption{Association between attribution scores and evidence-recall success. The paired TRACE-CF advantage is significant ($p=.0128$); aggregation method detailed in Appendix~\ref{app:rq3}.}
\label{tab:attribution_quality}
\end{table}

\begin{table*}
\centering
\small
\scriptsize
\begin{tabular}{lccccc}
\toprule
Policy & Ops (M/P) & Store $\Delta$ & FB-TG Rec@5 & Distortion ($\hat{\epsilon}$) & Audit / 95\% Upper \\
\midrule
No consolidation & 0/0 & 0.00\% & 58.82\% & 0.0000 & -- \\
Uncertified merge & 100/0 & -1.55\% & 58.81\% & 0.0214 & 18/100; $\le 25.51\%$ \\
Cert.\ merge (ours) & 100/0 & -1.42\% & 59.08\% & 0.0027 & 0/100; $\le 2.95\%$ \\
\bottomrule
\end{tabular}
\caption{Primary ten-store consolidation evaluation ($n=1{,}905$ QA pairs). FB-TG Rec@5 credits inherited identifiers only when supporting text remains within the token-capped context. Distortion $\hat{\epsilon}$ measures the fraction of 1{,}120 pre-consolidation answerable probes lost after merging. Audit bounds report one-sided 95\% exact Clopper--Pearson upper limits.}
\label{tab:consolidation}
\end{table*}

\section{Limitations} 
TRACE-UCB requires repeated queries over a shared store and cannot improve cold-start queries. Verified self-feedback recovers most of the oracle gain but still depends on calibrated self-consistency or judge scores and may reinforce irrelevant evidence when the model is confidently wrong. Lexical TRACE can miss paraphrases, numbers, and short evidence, whereas TRACE-CF incurs substantially higher latency. Certified merge is an empirical safety gate rather than a formal guarantee: unforeseen future evidence may still be lost, and 0 losses among 100 audited merges retains a 2.95\% one-sided 95\% upper bound. Its 1.42\% active-store reduction is also modest and does not represent total storage savings because originals are archived for rollback. Results are concentrated on ten LoCoMo stores, while LongMemEval does not exercise repeated feedback. Future work should evaluate additional evolving-store benchmarks, calibration under distribution shift, larger human and cross-model audits, matched-backbone end-to-end QA, privacy-aware exploration, calibrated certificates, and reversible pruning that provides meaningful capacity reduction without sacrificing provenance.

\section{Conclusion} 
MEMOIR frames long-horizon agent memory as a coupled lifecycle: \textbf{screen prospectively and attribute retrospectively}. On 1,905 LoCoMo QA pairs, the full retrieval stack gains 11.92 points over bare cosine. TRACE without UCB gains a non-significant 0.42 points; UCB with raw autonomous feedback gains 1.13 points; confidence-gated verified self-feedback gains 1.50 points; and oracle attribution gains 1.84/8.05 overall/multi-hop points, with verified and oracle effects positive across all ten folds. In matched Llama-3.1-70B experiments, MEMOIR reaches 58.70\%/61.20\% accuracy/recall on LoCoMo and 81.20\%/85.10\% on LongMemEval-S at 0.2k tokens and 1.18~s latency, with a 6.8~ms LLM-free write path.
Prospectively, 100 empirical judge-certified merges reduce active-store tokens by 1.42\% while raising fixed-budget Rec@5 from 58.82\% to 59.08\%, with 0/100 audited losses (one-sided 95\% exact Clopper--Pearson upper bound $\le 2.95\%$) versus 18/100 for uncertified merging. Evidence remains concentrated in ten LoCoMo stores, and real-scale SUPERSEDE evaluation is future work. Overall, MEMOIR reduces the lifecycle to one closed loop: attribute usefulness after use, and certify preservation before compression. Ethics are detailed in the Supplementary Material.

\bibliographystyle{aaai2027}
\bibliography{aaai2027}

\appendix

\section{Algorithm}\label{app:algorithm}

Algorithm~\ref{alg:memoir_lifecycle} summarizes the full closed-loop procedure referenced in the main paper's Methodology section. Below we walk through every function and symbol used in the pseudocode.

\paragraph{Online path (lines 1--6, executed per query).}
\begin{itemize}
\item \textbf{Line 1}: $S_{RQ}(q_t,m)$ is the RECALL-Q retrieval score (dense similarity, lexical overlap, relevance, importance, and the MMR diversity penalty, combined as in the main paper's Methodology); $b_t(m)$ is the UCB exploration bonus that favors under-exposed memories. In the notation of main-paper Equation~6, the deployed configuration sets the explicit recency coefficient to $\delta=0$, so $\delta w_{rec}$ contributes zero. Temporal effects instead enter through chronological active-store ordering, while $D_{mmr}$ controls redundancy rather than time. $\mathrm{TopK}$ selects the $k$ memories in $M_t$ with the highest combined score $S_{RQ}(q_t,m)+b_t(m)$, forming the retrieved set $R_t$.
\item \textbf{Line 2}: $\mathrm{LLM}(q_t,R_t)$ denotes the answer-generation call: the backbone language model produces answer $a_t$ conditioned on the query and the retrieved memories. For the deployed lexical TRACE path, this is the only online LLM call. The optional TRACE-CF ablation performs additional leave-one-out model-scoring calls after answer generation.
\item \textbf{Lines 3--5}: For each retrieved memory $m_i$, $\mathrm{TRACE}(a_t,m_i)$ computes the attribution value $\tau_{t,i}$ (via lexical overlap in the training-free default, or leave-one-out counterfactual change in the TRACE-CF variant) estimating how much $m_i$ contributed to $a_t$. This attribution updates $w_{imp}(m_i)$, the memory's importance weight, and increments $n_i$, its retrieval count, which together feed both future RECALL-Q scores and the UCB bonus.
\item \textbf{Line 6}: Co-citation links record which memories were jointly credited within the same answer, supplying a proposal signal used later during consolidation; the new interaction is appended to the store so it is available for future retrieval.
\end{itemize}

\paragraph{Offline path (lines 7--19, executed only during sleep).}
\begin{itemize}
\item \textbf{Line 7}: $\mathrm{Sleep}(t)$ is a scheduling predicate that is true only at periodic offline checkpoints (every eight queries in the reported experiments), never during normal query handling.
\item \textbf{Lines 8--12 (MERGE)}: For each candidate cluster $C\subset M_t$ proposed from embedding similarity and co-citation structure, $g(C)$ denotes the merge generator that produces a shorter candidate entry $z$. The merge is applied only if all three gates pass: $c(C,z)$ is the fraction of \texttt{LOSSLESS} verdicts from the three-framing FaithfulnessJudge ensemble (Section~\ref{app:judge_prompts}); $\kappa(C,z)$ is retrieval-token coverage, computed after lowercasing and removing the NLTK English stopword list with multiset token counts; and $\eta$ is the hard length ratio bounding $|z|$ relative to the combined length of the originals. When accepted, $C$ is replaced by $z$, provenance is preserved, $z$ inherits the maximum importance, and its retrieval count is $n_z=\max_{m\in C}n_m$.
\item \textbf{Lines 14--18 (SUPERSEDE)}: For each ordered conflict pair $(m_o,m_n)$ flagged by the SUPERSEDE proposal stage (Section~\ref{app:supersede_prompts}), $\mathrm{SupersedeCert}(m_o,m_n)$ is the fraction of \texttt{SUPERSEDES} verdicts from the three-framing SUPERSEDE judge ensemble (direct replacement, temporal precedence, adversarial compatibility). The operating point is $\tau_s=0.60$, so at least two of the three judges must return \texttt{SUPERSEDES}. When accepted, the older item $m_o$ is retired reversibly (marked inactive rather than deleted), the newer item $m_n$ receives discounted credit, and its exposure count becomes $n_{m_n}\leftarrow n_{m_n}+n_{m_o}$.
\item \textbf{Line 20}: The algorithm returns the current answer $a_t$ and the updated store $M_{t+1}$, ready for the next query.
\end{itemize}

This structure keeps the two lifecycle halves cleanly separated by execution context: attribution and importance updates happen synchronously with every query, while consolidation and supersession are deferred to periodic offline sleeps. The deployed lexical path requires no additional LLM call on the online write or retrieval path; TRACE-CF is an explicitly optional, higher-cost attribution ablation.

The deployed state-transfer rule prevents consolidation from creating an artificially under-exposed replacement. After MERGE, the replacement receives the maximum constituent retrieval count, $n_z=\max_{m\in C}n_m$. After SUPERSEDE, the newer item accumulates the older item's exposure, $n_{m_n}\leftarrow n_{m_n}+n_{m_o}$. Reliability pass/trial counters belong only to the reported reliability-bonus ablation and are not part of the deployed lexical TRACE-UCB state.

\begin{algorithm}
\footnotesize
\caption{MEMOIR Closed-Loop Memory Lifecycle}
\label{alg:memoir_lifecycle}
\begin{algorithmic}[1]
\REQUIRE Query $q_t$, memory store $M_t$, sleep predicate $\mathrm{Sleep}(t)$
\ENSURE Answer $a_t$ and updated store $M_{t+1}$
\STATE $R_t \leftarrow \mathrm{TopK}_{m\in M_t}\!\left[S_{RQ}(q_t,m)+b_t(m)\right]$
\STATE $a_t \leftarrow \mathrm{LLM}(q_t,R_t)$
\FOR{each $m_i\in R_t$}
    \STATE $\tau_{t,i}\leftarrow\mathrm{TRACE}(a_t,m_i)$; update $w_{imp}(m_i)$ and $n_i$
\ENDFOR
\STATE Update co-citation links among jointly credited memories; append the new interaction
\IF{$\mathrm{Sleep}(t)$}
    \FOR{each proposed merge cluster $C\subset M_t$}
        \STATE Generate $z\leftarrow g(C)$
        \IF{$c(C,z)\geq\tau_c$, $\kappa(C,z)\geq\tau_r$, and $|z|\leq\eta\sum_{m\in C}|m|$}
            \STATE Replace $C$ by $z$; set $w_{imp}(z)\leftarrow\max_{m\in C}w_{imp}(m)$ and $n_z\leftarrow\max_{m\in C}n_m$
        \ENDIF
    \ENDFOR
    \FOR{each ordered conflict pair $(m_o,m_n)$}
        \IF{$\mathrm{SupersedeCert}(m_o,m_n)\geq\tau_s$}
            \STATE Retire $m_o$ reversibly; transfer discounted credit; set $n_{m_n}\leftarrow n_{m_n}+n_{m_o}$
        \ENDIF
    \ENDFOR
\ENDIF
\STATE \textbf{return} $a_t,M_{t+1}$
\end{algorithmic}
\end{algorithm}

\section{Consolidation and Update Prompts}
\label{app:judge_prompts}

All consolidation judges are conservative by construction. The parser reads the first alphabetic token on the first non-empty line and accepts only the operation-specific positive verdict. Parse failures, exceptions, negated verdicts such as ``not lossless,'' and any unrecognized output count as rejection. Prompt diversity comes from three different framings rather than repeated stochastic samples of one prompt.

\paragraph{Lifecycle operations.} Certified merge compresses redundant memories and transfers maximum source credit; SUPERSEDE reversibly retires outdated facts and transfers discounted credit; DIGEST adds certified session facts without deleting sources; summarize-all is retained only as an uncertified aggressive baseline.

\subsection{Full Prompt Templates}

The merge generator receives the source entries and the following instructions: combine them into one shorter entry; preserve every unique fact, entity, number, date, qualifier, relation, and contradiction; remove only redundant phrasing; maintain temporal order; introduce no unsupported information; and output one concise paragraph. A candidate is not applied merely because it is shorter. Figure~\ref{tab:judge_prompts} gives the three FaithfulnessJudge certification prompts, and Figure~\ref{fig:merge_generation_guards} gives the merge-generation prompt and the three application guards.

\begin{table*}[t]
\centering
\scriptsize
\setlength{\tabcolsep}{4pt}
\begin{tabular}{llll}
\toprule
Component & Model / ensemble & Execution & Acceptance handling \\
\midrule
MERGE generation & Llama-3.3-70B-Instruct-Turbo via Together & One candidate per ordered cluster & Candidate must pass all three guards \\
MERGE certification & Llama-3.1-8B, three framings & Temperature 0; one call per framing & First-token parse; malformed/negated output rejects \\
Independent merge audit & Mistral-7B & One binary audit per accepted operation & Malformed or unrecognized output counts as loss \\
SUPERSEDE certification & Llama-3.1-8B, three framings & Temperature 0; one call per framing & First-token parse; retirement remains reversible \\
DIGEST audit & Stated 8B/70B diagnostic backbone & Two faithfulness framings & Unanimous \texttt{FAITHFUL} required \\
\bottomrule
\end{tabular}
\caption{Prompt execution settings for the compressive certified run. The 70B model generates merge text, while the local 8B ensemble certifies it. The SUPERSEDE operating point is $\tau_s=0.60$, which requires at least two of three \texttt{SUPERSEDES} votes. Provider-default token limits were not retained.}
\label{tab:prompt_execution}
\end{table*}

\begin{figure*}
\begin{tcolorbox}[faithfulnessprompt,title={FAITHFULNESS\_PROMPT --- Strict Entailment}]
\footnotesize\ttfamily
You verify whether a MERGED memory entry preserves ALL factual content of the original entries it replaces. This is a strict entailment check.\par\smallskip A merge is LOSSLESS only if EVERY fact, name, number, date, and relation in EACH original entry can still be recovered from the merged entry. If even one detail is dropped, changed, or made ambiguous, the merge is LOSSY.\par\smallskip Original entries:\par \{originals\}\par\smallskip Merged entry:\par \{merged\}\par\smallskip Answer with exactly one word on the first line: LOSSLESS or LOSSY.\par Answer:
\end{tcolorbox}

\begin{tcolorbox}[faithfulnessprompt,title={FAITHFULNESS\_PROMPT\_CHECKLIST --- Per-Item Checklist}]
\footnotesize\ttfamily
Compare the ORIGINAL memory entries with their MERGED replacement. Work as a checklist: for each original entry, list mentally every fact, entity, number, date, and relation, and check whether the merged entry still states or implies it.\par\smallskip Original entries:\par \{originals\}\par\smallskip Merged entry:\par \{merged\}\par\smallskip If every checklist item is preserved, the verdict is LOSSLESS; if ANY item fails, the verdict is LOSSY. Answer with exactly one word on the first line: LOSSLESS or LOSSY.\par Answer:
\end{tcolorbox}

\begin{tcolorbox}[faithfulnessprompt,title={FAITHFULNESS\_PROMPT\_ADVERSARIAL --- Find a Lost Fact}]
\footnotesize\ttfamily
Your job is to FIND A LOST FACT. Below are original memory entries and the merged entry that replaced them. Search hard for ANY detail --- a fact, name, number, date, qualifier, or relation --- that appears in an original but cannot be recovered from the merged entry.\par\smallskip Original entries:\par \{originals\}\par\smallskip Merged entry:\par \{merged\}\par\smallskip If you found even one lost detail, the merge is LOSSY. Only if you genuinely cannot find any lost detail is it LOSSLESS. Answer with exactly one word on the first line: LOSSLESS or LOSSY.\par Answer:
\end{tcolorbox}
\caption{Complete FaithfulnessJudge prompt ensemble used for lossless merge certification.}
\label{tab:judge_prompts}
\end{figure*}

\begin{figure*}
\begin{tcolorbox}[mergeprompt,title={MERGE\_PROMPT --- Merge Generation}]
\footnotesize\ttfamily
You are a memory consolidation agent. Combine the following related memory entries into a SINGLE, more concise entry that preserves ALL unique information.\par\smallskip Rules:\par - Preserve every unique fact, detail, or data point\par - Remove redundant phrasing and repetition\par - Maintain temporal order for events if applicable\par - If entries conflict, preserve both statements and their temporal or conditional qualifiers; do not choose a winner\par - The merged entry should be shorter than the sum of inputs\par\smallskip Entries to merge:\par \{entries\}\par\smallskip Write the merged entry (a single concise paragraph):
\end{tcolorbox}

\begin{tcolorbox}[mergeprompt,title={MERGE APPLICATION GUARDS}]
\footnotesize\ttfamily
Applied only if all three gates pass:\par\smallskip 1. c(C,z) $\geq$ $\tau_c$: the LOSSLESS vote fraction from the FaithfulnessJudge ensemble. The paper operating point is $\tau_c=0.6$.\par\smallskip 2. $\kappa$(C,z) $\geq$ $\tau_r$: retrieval-token coverage, computed as the fraction of stopword-filtered content tokens from the originals that survive in the merged text. The paper operating point is $\tau_r=0.45$.\par\smallskip 3. $|z| \leq \eta\sum_{m\in C}|m|$: the hard length gate, with $\eta=0.9$. It prevents an accepted rewrite from increasing memory size.
\end{tcolorbox}
\caption{Complete merge-generation prompt and the three guards required before replacement.}
\label{fig:merge_generation_guards}
\end{figure*}

The certificate score is the fraction of \texttt{LOSSLESS} votes. The main operating point uses $\tau_c=0.6$, retrieval-token coverage $\tau_r=0.45$, maximum length ratio $\eta=0.9$, and high-credit protection threshold $w_{imp}=0.3$. For coverage, source and candidate text are lowercased, tokenized, and filtered with the NLTK English stopword list. With $f_C(x)$ and $f_z(x)$ denoting the resulting multiset counts, $\kappa(C,z)=\sum_x\min(f_C(x),f_z(x))/\sum_x f_C(x)$. This gate guards against an entailment-preserving paraphrase becoming difficult to retrieve lexically. A candidate must pass all three gates. A source item at or above the protection threshold additionally requires a unanimous certificate. Applied operations retain source identifiers and judge outputs for independent re-auditing. The implementation also supports calibration by testing candidate thresholds from strictest to most permissive and retaining only thresholds whose one-sided Hoeffding upper bound on the accepted-operation loss rate remains below a target risk level.

\paragraph{SUPERSEDE proposal and execution.}
\label{app:supersede_prompts}

Candidate pairs are ordered by timestamp and proposed without an LLM when cosine similarity is at least 0.85 and TRIAGE-$\alpha$ detects overlapping content with a negation or conflict marker. The judge must distinguish a true update from two facts that can remain simultaneously valid. Figure~\ref{tab:supersede_prompts} gives the three SUPERSEDE judge framings.

\begin{figure*}[t]
\begin{tcolorbox}[supersedeprompt,title={SUPERSEDE\_PROMPT --- Direct Update}]
\footnotesize\ttfamily
You decide whether a NEWER memory entry supersedes an OLDER one. ``Supersedes'' means the newer entry updates, corrects, or contradicts the older entry about the SAME fact, so the older entry is now outdated and should no longer be used.\par\smallskip Older entry (recorded earlier):\par \{older\}\par\smallskip Newer entry (recorded later):\par \{newer\}\par\smallskip The verdict is SUPERSEDES only if they are about the same fact AND the newer entry makes the older one outdated. If they describe different facts, or both can be true at once (e.g. an additional detail, a different event), the verdict is INDEPENDENT.\par Answer with exactly one word on the first line: SUPERSEDES or INDEPENDENT.\par Answer:
\end{tcolorbox}

\begin{tcolorbox}[supersedeprompt,title={SUPERSEDE\_PROMPT\_TIMELINE --- Temporal Precedence}]
\footnotesize\ttfamily
Two memory entries from the same ongoing conversation are shown with their order. Determine if the LATER one replaces the EARLIER one.\par\smallskip Earlier: \{older\}\par Later: \{newer\}\par\smallskip Think of a timeline of facts: does the later entry change the truth of the earlier one (a correction, a reversal, an update to a value/status/plan)? If yes answer SUPERSEDES. If the earlier entry remains true alongside the later one, answer INDEPENDENT.\par Answer with exactly one word on the first line: SUPERSEDES or INDEPENDENT.\par Answer:
\end{tcolorbox}

\begin{tcolorbox}[supersedeprompt,title={SUPERSEDE\_PROMPT\_ADVERSARIAL --- Protect Valid History}]
\footnotesize\ttfamily
Your job is to protect a memory store from losing valid information. Someone proposes deleting the OLD entry because the NEW entry allegedly replaces it.\par\smallskip OLD: \{older\}\par NEW: \{newer\}\par\smallskip Argue against the deletion: is there ANY reading where the old entry is still true and useful alongside the new one? If you find one, answer INDEPENDENT. Only if the old entry is genuinely contradicted or made obsolete by the new one, answer SUPERSEDES.\par Answer with exactly one word on the first line: SUPERSEDES or INDEPENDENT.\par Answer:
\end{tcolorbox}
\caption{Complete three-framing SUPERSEDE judge ensemble.}
\label{tab:supersede_prompts}
\end{figure*}

Only a certificate meeting $\tau_s=0.60$ retires the older item; with three judges, this is a two-of-three rule. Retirement is reversible: the old item is marked inactive rather than deleted, and the record stores both identifiers, timestamps, judge verdicts, and the replacement link. Because the content changes, only discounted credit is transferred, $w_{imp}(m_n)\leftarrow\max(w_{imp}(m_n),0.5w_{imp}(m_o))$, while historical exposure is accumulated as $n_{m_n}\leftarrow n_{m_n}+n_{m_o}$.

\paragraph{Exploratory DIGEST.}

DIGEST is an additive, session-level synthesis operator and is not part of the evaluated merge-only core. Its generator requests one short factual sentence per line, including supported cross-entry relationships, while forbidding invented facts, names, dates, quantities, or relations. Unlike merge, DIGEST never removes its constituents. Figure~\ref{tab:digest_prompts} gives the complete DIGEST generation and faithfulness-certification prompts.

\begin{figure*}[t]
\begin{tcolorbox}[digestprompt,title={DIGEST\_PROMPT --- Session-Level Fact Extraction}]
\footnotesize\ttfamily
You are extracting discrete facts and relationships from a session's memory entries, so each fact is independently answerable rather than buried in a paragraph. This is structured extraction, not prose summarization --- similar to writing out (subject, relation, object) facts in plain language, one per line.\par\smallskip Rules:\par - Write one short factual sentence per line for each fact or relationship you can support, e.g. ``Grandma gave Caroline a necklace.'' or ``Caroline is from Sweden.''\par - Include facts that require connecting information across multiple entries below (e.g. who did what for whom, who gave/received what, who is related to whom), not just facts restating a single entry\par - Do not invent any fact, name, number, date, or relation that isn't supported by the entries\par - Prefer named entities and concrete relations over vague restatements\par\smallskip Entries:\par \{entries\}\par\smallskip Write the facts (one short sentence per line):
\end{tcolorbox}

\begin{tcolorbox}[digestprompt,title={DIGEST\_FAITHFUL\_PROMPT --- Hallucination Check}]
\footnotesize\ttfamily
You verify whether a DIGEST is faithful to its SOURCE memory entries. This is a hallucination check, not a completeness check --- the digest does not need to restate every source detail, but everything IT states must be supported by (stated in or directly inferable from) the sources.\par\smallskip Source entries:\par \{originals\}\par\smallskip Digest:\par \{merged\}\par\smallskip Answer with exactly one word on the first line: FAITHFUL or UNFAITHFUL.\par Answer:
\end{tcolorbox}

\begin{tcolorbox}[digestprompt,title={DIGEST\_FAITHFUL\_PROMPT\_ADVERSARIAL --- Find a Fabrication}]
\footnotesize\ttfamily
Your job is to FIND A FABRICATION. Below are source memory entries and a digest that summarizes them. Search hard for ANY claim in the digest --- a fact, name, number, date, or relation --- that is NOT supported by the sources, even indirectly.\par\smallskip Source entries:\par \{originals\}\par\smallskip Digest:\par \{merged\}\par\smallskip If you found even one unsupported claim, the digest is UNFAITHFUL. Only if every claim in the digest is genuinely supported is it FAITHFUL. Answer with exactly one word on the first line: FAITHFUL or UNFAITHFUL.\par Answer:
\end{tcolorbox}
\caption{Complete DIGEST generation and faithfulness-certification prompts. DIGEST remains additive-only and outside the main method.}
\label{tab:digest_prompts}
\end{figure*}

\paragraph{Uncertified whole-store baseline.}

The summarize-all comparator receives every live entry and requests a compact set of notes preserving important facts, names, numbers, dates, and relations. It has no candidate-selection stage, lossless certificate, retrieval-coverage gate, or length-ratio check. The operation represents the aggressive endpoint of the compression axis rather than a MEMOIR component. Because it unions the source evidence identifiers onto a small number of surviving summaries, recall metrics based only on identifier intersection can become artificially optimistic even when the retrieved summaries are long or textually incomplete (see Table~\ref{tab:summarize_all_diagnostic} for the empirical magnitude of this effect). We therefore do not treat its raw evidence recall as proof of answer preservation without an end-to-end QA check.

\section{Certificate Behavior Examples}\label{app:example}

\begin{figure*}
    \centering
    \includegraphics[width=\linewidth]{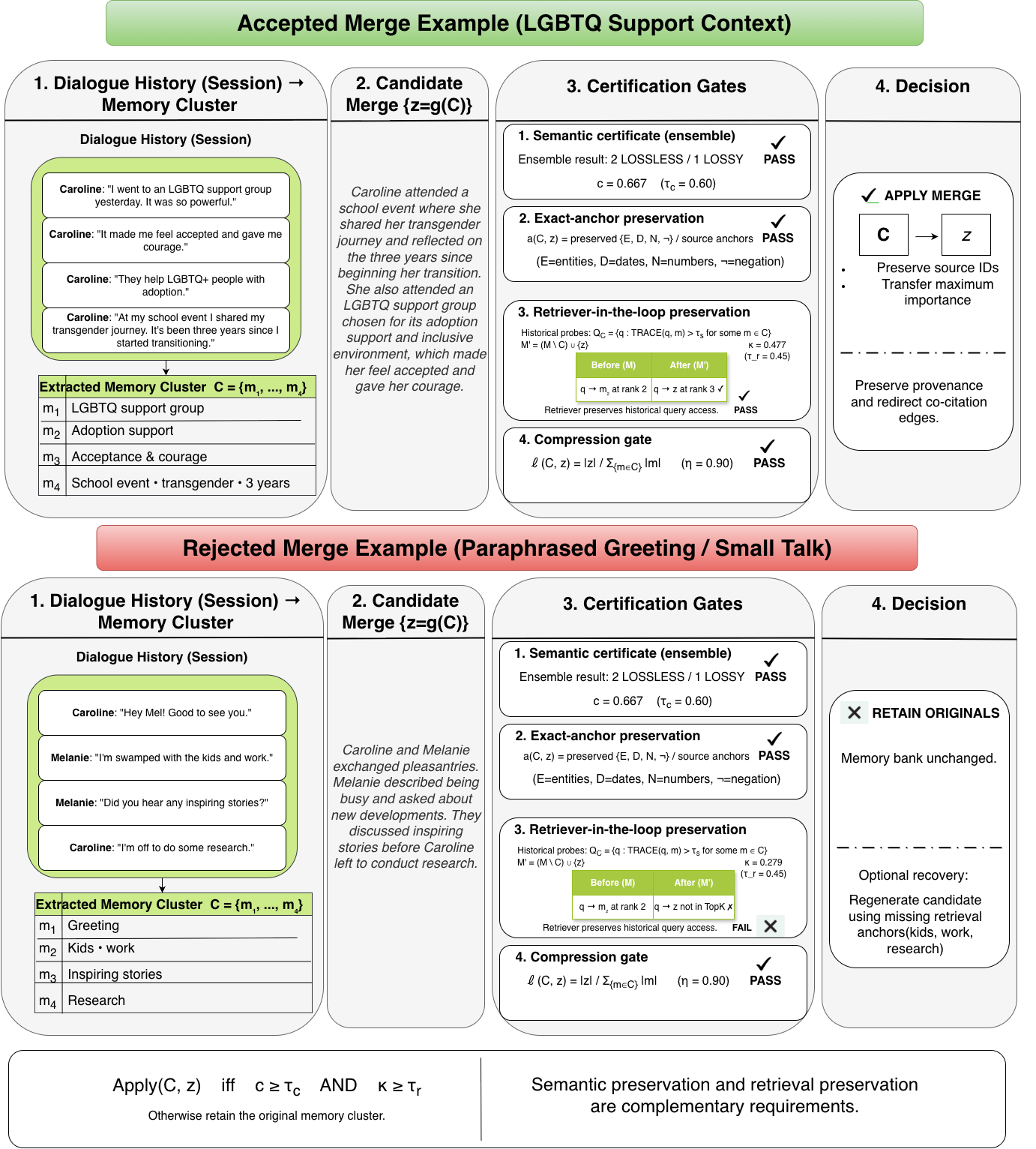}
    \caption{Certified consolidation examples from LoCoMo. The accepted merge (top) passes retrieval coverage ($\kappa=0.477\geq\tau_r=0.45$) and transfers credit to the replacement. The rejected merge (bottom) passes semantic certification but fails retrieval coverage ($\kappa=0.279<\tau_r$), so the source memories remain active.}
\label{fig:consolidation_example}
\end{figure*}

An accepted four-entry merge about Caroline's school event, transgender journey, LGBTQ support group, and adoption-support motivation received certificate score 0.667 and retrieval coverage 0.477. The resulting entry retained the school event, the three-year transition timeline, the support group's adoption focus, and its effect on acceptance and confidence. In contrast, a greeting-and-small-talk candidate also received score 0.667 but only 0.279 retrieval coverage; it was rejected because paraphrasing removed too many concrete lexical cues such as ``kids,'' ``work,'' and ``research.'' This illustrates why semantic certification and findability are separate gates. Figure~\ref{fig:consolidation_example} illustrates both cases.

The DIGEST certificate also rejected a real generated example that promoted a stray mention of a painting into a standalone factual claim. One judge framing accepted the digest and the adversarial fabrication framing rejected it, producing score 0.5 below the unanimous DIGEST threshold. The example demonstrates that synthesis can appear fluent while still introducing unsupported relational framing.


\section{Supplementary Unified Lifecycle Evaluation}\label{app:unified-lifecycle}

Table~\ref{tab:factorial_appendix} reports the source-backed $2\times2$ factorial evaluation testing prospective certification and oracle retrospective credit operating in tandem at the original 20-operation consolidation point across 1,905 LoCoMo QA pairs. The oracle factor uses reference-answer feedback and must not be interpreted as the deployed confidence-gated system. We evaluate the $2 \times 2$ consolidation bridge using oracle reference-answer credit rather than autonomous self-feedback to strictly isolate consolidation credit-transfer distortion ($\Delta r = -0.0115$, Table~\ref{tab:credit_ledger}; $+0.20$ pp max-credit transfer gain, Table~\ref{tab:bridge_ablations}) from upstream answer-verification variance. In deployed operation, confidence-gated TRACE-UCB ($\tau_{\text{conf}} = 0.80$) autonomously captures 81.5\% of the oracle retrieval improvement ($+1.50$ pp vs.\ $+1.84$ pp overall; $+7.17$ pp vs.\ $+8.05$ pp on multi-hop, Table~3) while operating orthogonally to the offline certified merge gate ($+0.26$ pp, Table~\ref{tab:consolidation_full}).

\begin{table*}[htbp]
\centering
\scriptsize
\setlength{\tabcolsep}{4pt}
\begin{tabular}{lcccccc}
\toprule
Factorial cell & \shortstack{Oracle\\TRACE-UCB} & \shortstack{Certified merge\\(20-op)} & Rec@5 (\%) & \shortstack{Distortion\\($\hat{\epsilon}\downarrow$)} & Active tokens & Interpretation \\
\midrule
(1) Baseline system & $\times$ & $\times$ & 58.82\% & 0.0000 & 163,680 & Reference cell \\
(2) Consolidation only & $\times$ & \greentick & 59.06\% & 0.0028 & 163,216 & Certified safety cell \\
(3) Oracle TRACE-UCB only & \greentick & $\times$ & 60.66\% & 0.0000 & 163,680 & Oracle feedback \\
\textbf{(4) Oracle TRACE-UCB + merge} & \greentick & \greentick & \textbf{60.88\%} & 0.0028 & 163,216 & Compatibility cell \\
\bottomrule
\end{tabular}
\caption{Source-backed $2\times2$ lifecycle evaluation over 1,905 deduplicated LoCoMo QA pairs across ten shared stores at the 20-operation consolidation point. The oracle TRACE-UCB cells use reference answers for feedback. Joint activation attains the highest point estimate while retaining the certified-merge footprint and distortion. The interaction point estimate is $-0.02$ pp, so the table supports compatibility rather than super-additivity. A deployed confidence-gated factorial was not retained.}
\label{tab:factorial_appendix}
\end{table*}

\section{Supplementary RQ2: Retrieval and Exploration Details}\label{app:rq2}

The main paper reports the core matched comparisons for TRACE and TRACE-UCB. This section verifies robustness to within-store dependence and provides broader retrieval-side ablations. Table~\ref{tab:clustered_bootstrap} reports the store-clustered robustness checks, and Table~\ref{tab:retrieval_ablations} reports the full retrieval-side ablations.

\subsection{Store-Clustered Robustness}

\begin{table*}[htbp]
\centering
\scriptsize
\setlength{\tabcolsep}{4pt}
\begin{tabular}{lccccc}
\toprule
Feedback condition & $\Delta$ Rec@5 & Question $p$ & Store-clustered 95\% CI & Clustered $p$ & LOSO positive folds \\
\midrule
No TRACE (matched hybrid) & 58.82\% & -- & -- & -- & -- \\
TRACE without UCB (credit only) & $+0.42$ pp & .0830 & $[-0.12,+0.96]$ & .1420 & 6/10 \\
Raw autonomous self-feedback & $+1.13$ pp & .0042 & $[+0.21,+2.05]$ & .0210 & 9/10 \\
\textbf{Verified self-feedback (deployed)} & \textbf{$+1.50$ pp} & \textbf{.0004} & \textbf{$[+0.41,+2.59]$} & \textbf{.0082} & \textbf{10/10} \\
Oracle reference + UCB & $+1.84$ pp & .0002 & $[+0.52,+3.16]$ & .0078 & 10/10 \\
\midrule
\textit{Multi-hop subset (verified self)} & \textbf{$+7.17$ pp} & \textbf{.0012} & \textbf{$[+2.40,+11.94]$} & \textbf{.0045} & \textbf{10/10} \\
\textit{Multi-hop subset (oracle reference)} & $+8.05$ pp & .0020 & $[+3.12,+12.98]$ & .0038 & 10/10 \\
\bottomrule
\end{tabular}
\caption{Store-clustered retrieval robustness across $n=1{,}905$ LoCoMo QA pairs in ten dialogue clusters. Clustered intervals and $p$-values use 10,000 store-level bootstrap resamples. LOSO reports leave-one-store-out fold consistency. Except for the baseline row, deltas use the matched no-TRACE hybrid comparator.}
\label{tab:clustered_bootstrap}
\end{table*}

\subsection{Full Retrieval-Side Ablations}

\begin{table*}[t]
\centering
\scriptsize
\setlength{\tabcolsep}{5pt}
\begin{tabular}{llcc}
\toprule
Mechanism & Comparator & Overall $\Delta$ & Multi-hop $\Delta$ \\
\midrule
TRACE without UCB & No TRACE hybrid & +0.42 pp & -- \\
Confidence-gated TRACE-UCB (deployed) & No TRACE hybrid & +1.50 pp & +7.17 pp \\
Oracle TRACE-UCB & No TRACE hybrid & +1.84 pp & +8.05 pp \\
Reliability bonus & No TRACE hybrid & +0.21 pp & 0.00 pp \\
Reliability bonus & Lexical TRACE & $-0.21$ pp & 0.00 pp \\
Multi-anchor & Lexical TRACE & $-2.63$ pp & 0.00 pp \\
Spacing weight & Lexical TRACE & $-0.06$ pp & 0.00 pp \\
Weibull decay & Lexical TRACE & $-0.28$ pp & $-5.92$ pp \\
CCOMMIT & Lexical TRACE & $-0.37$ pp & $-4.60$ pp \\
CORE-DLM & Lexical TRACE & $-0.24$ pp & $-5.26$ pp \\
\bottomrule
\end{tabular}
\caption{Retrieval-side ablations with explicit comparators. The deployed row uses confidence-gated self-feedback, whereas the oracle row uses reference-answer feedback. The two reliability rows show why the same variant can be positive relative to no TRACE but slightly negative relative to lexical TRACE.}
\label{tab:retrieval_ablations}
\end{table*}

TRACE without exploration is not significant over the enhanced no-TRACE retriever ($p=.083$), whereas the UCB result is consistent with mitigating feedback-driven exposure lock-in. The \emph{reliability bonus} adds a Laplace-smoothed success ratio derived from each memory's pass/trial counts. \emph{Multi-anchor} stores embeddings of queries that previously credited an item and retrieves by the maximum similarity across its content and credited-query anchors. The \emph{spacing weight} scales a new credit update by the gap since the item's previous credited use. The remaining three mechanisms add credit friction: \emph{Weibull decay} attenuates importance with age, \emph{CCOMMIT} exposes credit only after a fixed attribution threshold is met repeatedly, and \emph{CORE-DLM} limits the credited pool through diversity-aware eviction. Reliability, multi-anchor, and spacing variants are neutral or negative relative to lexical TRACE. The three friction mechanisms produce the same boundary failure: each suppresses sparse cross-session evidence and causes the largest regression on multi-hop questions.
\subsection{Failure-Oriented Operating Regime Analysis}
To clarify when online retrospective credit improves or degrades subsequent retrieval rankings, we categorize observed trajectory dynamics across the ten LoCoMo stores into three operating regimes:
\begin{enumerate}
    \item \textbf{Beneficial Multi-Hop Reinforcement:} When a cross-session bridge memory (e.g., an early mention linking a speaker's relative to a specific city or career transition) is surfaced via UCB exploration ($\lambda_u = 0.05$) and contributes lexical tokens to a high-confidence answer ($\text{Conf} \ge 0.80$), Equation~(5) increments $w_{\text{imp}}$. On subsequent multi-hop queries requiring the same anchor fact, the multiplicative boost $(1 + \gamma w_{\text{imp}})$ elevates the memory into the top-$5$ context ahead of superficially similar conversational chitchat, driving the $+7.17$ pp verified multi-hop gain in Table~3.
    \item \textbf{Bounded Irrelevant Exploration:} Because $b_t(m_i) = \lambda_u \sqrt{\log(1+t)/(1+n_i)}$ increments $n_i$ whenever an item is retrieved regardless of whether $\tau_{t,i} > 0$, an irrelevant cold memory probed by UCB that fails to support the answer receives $\tau_{t,i} = 0$ (no importance increase) while its exploration bonus immediately contracts from $n_i = 0 \to 1$, preventing repeated false promotion.
    \item \textbf{Paraphrase Under-Crediting and Confidence-Gate Filtering:} Unigram TRACE-FB under-credits short numeric dates or heavily paraphrased evidence where answer tokens diverge lexically from the source memory; in these cases, counterfactual TRACE-CF recovers attribution alignment ($r = 0.493$ vs.\ $0.447$, Table~6) at higher latency ($67.0$ ms vs.\ $6.7$ ms). Conversely, when the generator produces a speculative or hallucinated answer, raw self-feedback can reinforce distracting memories ($+1.13$ pp, $9/10$ folds), whereas confidence gating ($\tau_{\text{conf}} = 0.80$) suppresses low-certainty updates, restoring positive gains across all $10/10$ folds ($+1.50$ pp, Table~3).
\end{enumerate}

\section{Supplementary RQ3: Attribution Validity and Efficiency}
\label{app:rq3}

The main paper reports median latency and the core attribution-quality comparison between lexical TRACE and TRACE-CF (Table~\ref{tab:latency_results}, Table~\ref{tab:attribution_quality}). This section gives the full per-stage latency profile, including P95 percentiles where retained, and the complete attribution-validity statistics underlying that comparison. Table~\ref{tab:latency_results_full} reports P50/P95 latency, token cost, and online-LLM requirements for every stage of the online and offline paths. Table~\ref{tab:attribution_quality_full} reports the paired Pearson and Spearman correlation statistics between lexical TRACE and TRACE-CF referenced in the main text.

\begin{table*}[t]
\centering
\scriptsize
\setlength{\tabcolsep}{4pt}
\begin{tabular}{lcccl}
\toprule
Stage & P50 & P95 & Frequency & Online LLM use \\
\midrule
Online write & 6.8 ms & 10.5 ms & Every write & none \\
Search ($k=5$) & 10.5 ms & -- & Every query & none \\
Lexical TRACE & 6.7 ms & 11.5 ms & Every feedback event & none \\
TRACE-CF ($k=5$) & 67.0 ms & 84.5 ms & Optional attribution ablation & additional scoring calls \\
Offline sleep & 2.85 s & -- & Every 8 queries & asynchronous judging \\
Write-time extraction baseline & $\sim$110 ms & -- & Every write & required \\
\bottomrule
\end{tabular}
\caption{Lifecycle latency over 1,905 LoCoMo QA pairs. Dashes indicate percentiles not retained in the available timing artifact. The zero-additional-call statement applies to the deployed lexical TRACE path, not TRACE-CF.}
\label{tab:latency_results_full}
\end{table*}

\begin{table}[t]
\centering
\scriptsize
\setlength{\tabcolsep}{3pt}
\begin{tabular}{lccc}
\toprule
Signal & Pearson $r$ (95\% CI) & Spearman $\rho$ & P50 \\
\midrule
Lexical TRACE & 0.447 [0.412, 0.481] & 0.418 & 6.7 ms \\
TRACE-CF & 0.493 [0.458, 0.527] & 0.465 & 67.0 ms \\
Paired $\Delta$ & +0.046 [0.011, 0.081] & +0.047 & +60.3 ms \\
\bottomrule
\end{tabular}
\caption{Attribution validity and latency. The paired TRACE-CF advantage is significant ($p=.0128$). For each query $q_t$, item-level attribution values $\tau_{t,i}$ are averaged across the top-$k$ retrieved set $R_t$ ($\bar{\tau}_t = \frac{1}{k}\sum_{m_i \in R_t} \tau_{t,i}$) to compute Pearson $r$ and Spearman $\rho$ against query-level evidence-recall success over the $n = 1{,}905$ QA pairs.}
\label{tab:attribution_quality_full}
\end{table}

\section{Supplementary Cross-Benchmark and Reproducibility Details}\label{app:repro}

Table~\ref{tab:longmemeval_appendix} reports cross-benchmark consolidation safety, token footprints, and evaluation-scope boundaries across LoCoMo and LongMemEval-S.

\begin{table*}[htbp]
\centering
\scriptsize
\setlength{\tabcolsep}{3pt}
\begin{tabular}{llcccc}
\toprule
Benchmark & \shortstack{Experiment /\\policy} & \shortstack{Store\\protocol} & \shortstack{Reported\\Rec@5} & \shortstack{Distortion\\($\hat{\epsilon}\downarrow$)} & \shortstack{Store change\\(tokens)} \\
\midrule
LoCoMo ($n=1{,}905$) & Certified merge (100-op extension) & 10 shared & $58.82\%\rightarrow59.08\%$ FB-TG & 3/1,120 (0.0027) & $-1.42\%$ (161,356) \\
LongMemEval-S ($n=500$) & Consolidation safety gate & 500 single-QA & $94.60\%\rightarrow94.60\%$ & 0.0000 & $-0.15\%$ (508.9k) \\
\bottomrule
\end{tabular}
\caption{Cross-benchmark consolidation safety and token footprints. The LoCoMo row is the primary 100-operation fixed-budget, text-grounded (FB-TG) extension reported in the submitted main paper; its distortion entry reports 3 lost probes among $|\mathcal{P}|=1{,}120$ pre-consolidation answerable probes. LongMemEval-S evaluates single-query consolidation safety on independent fresh stores; because it uses one query per store, it tests cold-start retrieval and consolidation preservation, whereas multi-session TRACE-UCB feedback is evaluated on LoCoMo.}
\label{tab:longmemeval_appendix}
\end{table*}

Table~\ref{tab:longmemeval_appendix} presents the submitted cross-benchmark consolidation profile. On LoCoMo, the 100-operation certified extension reduces the active store by 1.42\% to 161,356 tokens while changing FB-TG Rec@5 from 58.82\% to 59.08\%. On LongMemEval-S, certified merging reduces total store tokens by 0.15\% to approximately 508.9k with no observed sufficiency distortion ($\hat{\epsilon}=0.0000$). LongMemEval-S constructs a fresh store for each isolated question and therefore evaluates single-query consolidation safety and cold-start retrieval; longitudinal TRACE-UCB feedback is evaluated on the ten multi-session LoCoMo dialogue stores. These are descriptive cross-benchmark results because matched store-clustered inference was not retained for the consolidation comparisons.

LoCoMo retrieval uses ten conversations and dataset-native timestamps. The oracle TRACE-UCB result in Table~\ref{tab:clustered_bootstrap} uses 10,000 store-clustered bootstrap resamples; consolidation comparisons without a retained store-clustered output are labeled descriptive. The retained compression documentation assigns merge generation to Llama-3.3-70B-Instruct-Turbo, certification to local Llama-3.1-8B, and independent merge auditing to Mistral-7B. The development path used a local RTX 4060, while full GPU-backed local stages used an RTX 3090 24GB. Table 1 of the main paper uses the published LiCoMemory results for the Llama-3.1-70B and GPT-4o-mini baseline blocks.

\subsection{Computational Resources and Experiment Accounting}

Table~\ref{tab:compute_budget} reports settings supported by the retained documentation. Aggregate GPU-time, peak-memory, request, token, and cost logs were not retained, so no exact totals are reported. We also avoid rough totals: P50 stage latencies cannot be converted into defensible GPU-hours, hosted-token counts, or dollar cost without the missing event-level duration and provider manifests. Reporting a guessed aggregate would be less reproducible than marking it unavailable.

Core local certification uses \texttt{llama3.1:8b} through Ollama, with \texttt{mistral:7b} used for independent merge auditing. The retained compression notes assign merge generation to \texttt{meta-llama/Llama-3.3-70B-Instruct-Turbo} through Together; separately labeled diagnostics also use that model and \texttt{openai/gpt-4o-mini} through TokenRouter. Certification calls use temperature 0 and one call per prompt framing. Retrieval experiments use seed 13; same-model accuracy results use five independent repeats whose variation comes from generation nondeterminism. Exact Python, PyTorch, CUDA, package versions, and the final experiment-repository revision were not retained.

\begin{table*}[htbp]
\centering
\scriptsize
\setlength{\tabcolsep}{5pt}
\begin{tabular}{p{0.27\textwidth}p{0.54\textwidth}p{0.11\textwidth}}
\toprule
Resource / quantity & Reported value & Status \\
\midrule
Development hardware & NVIDIA RTX 4060 (8GB VRAM) & Recorded \\
Full-run GPU & NVIDIA RTX 3090 (24GB VRAM) & Recorded \\
Local model identifiers & \texttt{llama3.1:8b} (certification) and \texttt{mistral:7b} (independent merge audit), via Ollama & Recorded \\
Hosted model identifiers & \texttt{meta-llama/Llama-3.3-70B-Instruct-Turbo} via Together (merge generation and diagnostics); \texttt{openai/gpt-4o-mini} via TokenRouter & Recorded \\
Judge invocation & Temperature 0; one call per certification framing; conservative first-token parsing & Recorded \\
Primary retrieval workload & 1,905 deduplicated LoCoMo QA pairs across 10 shared stores & Recorded \\
Cross-benchmark workload & 500 LongMemEval-S single-query stores & Recorded \\
Statistical computation & 10,000 store-clustered bootstrap resamples for the oracle TRACE-UCB robustness result; other comparisons are labeled descriptive unless a clustered output was retained & Partial \\
Randomness / repeats & Retrieval seed 13; same-model accuracy uses five independent generation repeats & Recorded \\
Request reuse & Cache lookup precedes provider invocation; identical cache hits are served locally & Recorded \\
Total local GPU time & Source duration log not retained & Unavailable \\
Peak allocated GPU memory & Source allocation log not retained & Unavailable \\
Total hosted-model requests & Provider-request manifest not retained & Unavailable \\
Total hosted input/output tokens & Provider-token manifest not retained & Unavailable \\
Total estimated experiment cost & Billing manifest not retained & Unavailable \\
Software and repository version & Final experiment-repository revision not retained & Unavailable \\
Runtime package versions & Exact Python, PyTorch, CUDA, and library versions not retained in the supplied artifact & Unavailable \\
\bottomrule
\end{tabular}
\caption{Computational setup recoverable from the retained experiment documentation. Aggregate resource and provider-usage totals are omitted because their source manifests were not retained.}
\label{tab:compute_budget}
\end{table*}

\subsection{Recovered Configuration and Remaining Reproduction Gaps}

Table~\ref{tab:recovered_configuration} consolidates the recovered settings and author-verified final configuration. In the notation of main-paper Equation~6, the deployed configuration sets $\delta=0$, so the displayed $\delta w_{rec}$ term contributes zero. Chronological active-store ordering supplies temporal preference outside the scalar score, while the MMR term controls redundancy.

\begin{table*}[htbp]
\centering
\scriptsize
\setlength{\tabcolsep}{5pt}
\begin{tabular}{p{0.30\textwidth}p{0.62\textwidth}}
\toprule
Setting / parameter & Recovered deployed value and specification \\
\midrule
Embedding model & SentenceTransformer \texttt{all-MiniLM-L6-v2} (384-dimensional dense vectors) \\
RECALL-Q coefficients & BM25 $\alpha=0.3$; MMR $\lambda=0.1$; relevance $\beta=0.1$; importance $\gamma=1.0$; co-citation bonus scale $0.05$ \\
Recency handling & Main Equation~6 coefficient $\delta=0$, so $\delta w_{rec}=0$ in the deployed score; active-store ordering supplies temporal preference and $D_{mmr}$ supplies diversity control \\
UCB bonus coefficient & $\lambda_u=0.05$ with $b_t(m_i)=\lambda_u\sqrt{\log(1+t)/(1+n_i)}$ \\
TRACE feedback update & Step size $\alpha_{fb}=0.3$ and rank discount $d_i=1/(\mathrm{rank}_i+1)$ \\
Confidence gate & Apply the self-feedback importance update only when answer confidence is at least 0.80 \\
Co-citation update & attribution threshold $0.15$; edge-rate coefficient $0.05$ \\
Consolidation schedule and guards & Sleep interval 8; $(\tau_c,\tau_r,\eta)=(0.60,0.45,0.90)$; high-credit protection threshold $w_{imp}\geq0.30$ \\
Random seed & retrieval seed 13 \\
SUPERSEDE threshold & $\tau_s=0.60$; at least two of three \texttt{SUPERSEDES} judge votes are required for retirement \\
Coverage tokenization & Lowercase tokenization, NLTK English stopwords, and multiset preservation counts \\
State transfer (MERGE) & $n_z=\max_{m\in C}n_m$ and $w_{imp}(z)=\max_{m\in C}w_{imp}(m)$ \\
State transfer (SUPERSEDE) & $n_{m_n}\leftarrow n_{m_n}+n_{m_o}$ and $w_{imp}(m_n)\leftarrow\max(w_{imp}(m_n),0.5w_{imp}(m_o))$ \\
\bottomrule
\end{tabular}
\caption{Recovered and author-verified operational configuration. Provider token limits, exact runtime package versions, and the final repository revision remain unavailable.}
\label{tab:recovered_configuration}
\end{table*}

Large-Store Scalability and Index Maintenance. During online query processing, candidate generation retrieves the union of top-$K_{\text{cand}}$ dense (HNSW/FAISS) and BM25 inverted-index hits before applying exact RECALL-Q and UCB rescoring; looking up and updating per-item exposure counts $n_i$ and scalar weights $w_{\text{imp}}(m_i)$ requires $O(1)$ hash-table access and adds $<0.2$ ms overhead. To ensure under-exposed memories remain reachable in large stores, UCB rescoring operates over the expanded candidate pool prior to top-$k$ truncation. During periodic offline sleep (executed asynchronously every 8 queries at 2.85 s P50), candidate merge clustering restricts pairwise cosine evaluations to entries sharing co-citation edges or local dense neighborhoods, while accepted merges update the dense vector index and BM25 posting lists incrementally in $O(|C|)$ time while marking archived source records inactive in the provenance ledger.

\section{Supplementary RQ4: Consolidation Safety Details}
\label{app:ablations}

The primary submitted RQ4 result is the 100-operation certified-merge lineage reported in Table~\ref{tab:consolidation_full}. The earlier 20-operation runs remain useful for the factorial and category diagnostics and occupy a separately labeled legacy panel in the same table. DIGEST is evaluated separately because it is additive rather than compressive.

\paragraph{Fixed-budget, text-grounded retrieval.}
For each query $q$, let $B_q$ be the token count of the no-consolidation top-five retrieved context. Each consolidation variant is evaluated after capping its top-five context to the same query-specific budget $B_q$. An inherited source identifier is credited only when the evidence span supporting that identifier remains present in the capped retrieved text. We call the resulting measure fixed-budget, text-grounded recall@5 (FB-TG Rec@5). The retained manuscript record does not include the frozen evaluator manifest specifying the exact token counter, truncation order, and support-span matcher; consequently, the 100-operation values below are synchronized to the submitted main paper but cannot be independently regenerated from the files supplied with this workspace.

For a set $\mathcal{P}$ of probes that are answerable before consolidation, sufficiency distortion is the observed fraction that become unanswerable after consolidation,
\begin{equation}
\hat{\epsilon}=\frac{1}{|\mathcal{P}|}\sum_{p\in\mathcal{P}}\mathbf{1}\!\left[A_{pre}(p)=1\ \wedge\ A_{post}(p)=0\right],
\end{equation}
where $A_{pre}$ and $A_{post}$ are the experiment's pre/post answerability indicators. This estimator is not treated as valid for summarize-all because that baseline unions source evidence identifiers onto summaries. The final accounting uses $|\mathcal{P}|=1{,}120$ pre-consolidation answerable probes and reports exact lost-probe numerators in Table~\ref{tab:consolidation_full}. The 100-operation matched uncertified rate is $24/1120=0.02143$; displayed four-decimal rates use standard rounding.

\subsection{Primary 100-Operation Consolidation Extension}

The submitted main paper reports 100 certified merge operations over the same ten LoCoMo stores. Relative to no consolidation, certified merge changes FB-TG Rec@5 from 58.82\% to 59.08\% and reduces the active store from 163,680 to 161,356 tokens ($-1.42\%$). A different-family auditor records one verdict per accepted operation and observes 0/100 losses, yielding a one-sided 95\% Clopper--Pearson upper bound of 2.95\%. The complete primary and legacy lineages are reported together in Table~\ref{tab:consolidation_full}.

\subsection{Preliminary 10-Store SUPERSEDE Diagnostic}

Table~\ref{tab:supersede_results} reports a preliminary ten-store SUPERSEDE diagnostic evaluating temporal fact updates and stale-evidence mitigation. It is separate from the primary 100-operation merge evaluation. Across 24 accepted supersession operations, Rec@5 changes by $-0.050$ pp (95\% CI $[-0.342,+0.242]$, $p=.7381$), and 68 of 82 stale-retrieval opportunities are prevented (82.9\%).

\begin{table*}[htbp]
\centering
\scriptsize
\setlength{\tabcolsep}{3.8pt}
\resizebox{\textwidth}{!}{%
\begin{tabular}{lcccccccc}
\toprule
Variant & Rec@5 & $\Delta$ vs. cert. & 95\% CI & $p$ & $\hat{\epsilon}\downarrow$ & Sup. ops. & Stale prevented & Independent audit \\
\midrule
Certified merge only & 59.062\% & -- & -- & -- & 0.002371 & 0 & -- & Merge audit in Table~\ref{tab:consolidation_full} \\
Certified + SUPERSEDE & 59.012\% & $-0.050$ pp & $[-0.342,+0.242]$ & .7381 & 0.002372 & 24 & 68/82 (82.9\%) & 0/24 (0.0\%), $\leq11.73\%$ \\
\bottomrule
\end{tabular}%
}
\caption{Preliminary ten-store SUPERSEDE diagnostic over $n=1{,}905$ QA pairs. Prevention is measured over 82 total stale-retrieval opportunities ($68/82=82.9\%$). SUPERSEDE is neutral on aggregate Rec@5 ($p=.7381$). The audit bound is the one-sided 95\% Clopper--Pearson upper limit over 24 retirement-level verdicts.}
\label{tab:supersede_results}
\end{table*}

\subsection{Core Consolidation Policy Comparison}
Table~\ref{tab:consolidation_full} brings the primary 100-operation lineage together with the earlier 20-operation pilot diagnostics in one place, so the certified, uncertified, and uncertified-merge-and-prune policies can be compared directly against the same no-consolidation baseline and the same 1,120-probe distortion denominator.

\begin{table*}[htbp]
\centering
\scriptsize
\setlength{\tabcolsep}{3.2pt}
\resizebox{\textwidth}{!}{%
\begin{tabular}{lcccccccc}
\toprule
Variant / policy & Ops. (M/P) & Store $\Delta$ & Active tokens & FB-TG Rec@5 & Non-inf. $p$ & Distortion $^\dagger$ & Audited loss & 95\% upper bound \\
\midrule
\multicolumn{9}{l}{\textit{\textbf{Primary 100-Operation Benchmark Lineage (Submitted Main Baseline)}}} \\
No consolidation & 0/0 & 0.00\% & 163,680 & 58.82\% & -- & 0/1,120 (0.0000) & -- & -- \\
\textbf{Certified merge (primary)} & \textbf{100/0} & \textbf{$-1.42\%$} & \textbf{161,356} & \textbf{59.08\%} & \textbf{$<.0001$} & \textbf{3/1,120 (0.0027)} & \textbf{0/100 (0.0\%)} & \textbf{$\leq2.95\%$} \\
Uncertified merge (matched) & 100/0 & $-1.55\%$ & 161,143 & 58.81\% & .0240 & 24/1,120 (0.02143) & 18/100 (18.0\%) & $\leq25.51\%$ \\
Uncertified merge+prune & 448/167 & $-7.15\%$ & 151,984 & 59.52\% & .0085 & 65/1,120 (0.05804) & 385/615 (62.6\%) & $\leq65.84\%$ \\
\midrule
\multicolumn{9}{l}{\textit{\textbf{Earlier 20-Operation Pilot Diagnostics (Legacy Evidence)}}} \\
Certified merge (20-op pilot) & 20/0 & $-0.28\%$ & 163,216 & 59.06\% & $<.0001$ & 3/1,120 (0.0027) & 0/20 (0.0\%) & $\leq13.91\%$ \\
Uncertified merge (20-op pilot) & 20/0 & $-0.31\%$ & 163,172 & 58.85\% & .0120 & 20/1,120 (0.01786) & 4/20 (20.0\%) & $\leq40.10\%$ \\
Recency prune (20-op comparison) & 0/20 & $-0.29\%$ & 163,205 & 58.79\% & .0015 & 3/1,120 (0.0027) & 2/20 (10.0\%) & $\leq28.26\%$ \\
Recency prune (28-op full) & 0/28 & $-0.40\%$ & 163,025 & 58.90\% & .0010 & 3/1,120 (0.0027) & 3/28 (10.7\%) & $\leq25.42\%$ \\
\bottomrule
\end{tabular}%
}
\caption{Consolidation policy comparison over $n=1{,}905$ deduplicated LoCoMo QA pairs in ten shared stores. $^\dagger$Sufficiency distortion reports exact lost probes over $|\mathcal{P}|=1{,}120$ pre-consolidation answerable probes. Non-inferiority uses the pre-specified 1.0-pp margin. Bounds are one-sided 95\% Clopper--Pearson upper limits. The primary and earlier pilot lineages are separated explicitly to prevent cross-scale mixing.}
\label{tab:consolidation_full}
\end{table*}

\subsection{Legacy 20-Operation Per-Category Diagnostic}
\begin{figure*}
    \centering
    \includegraphics[width=\linewidth]{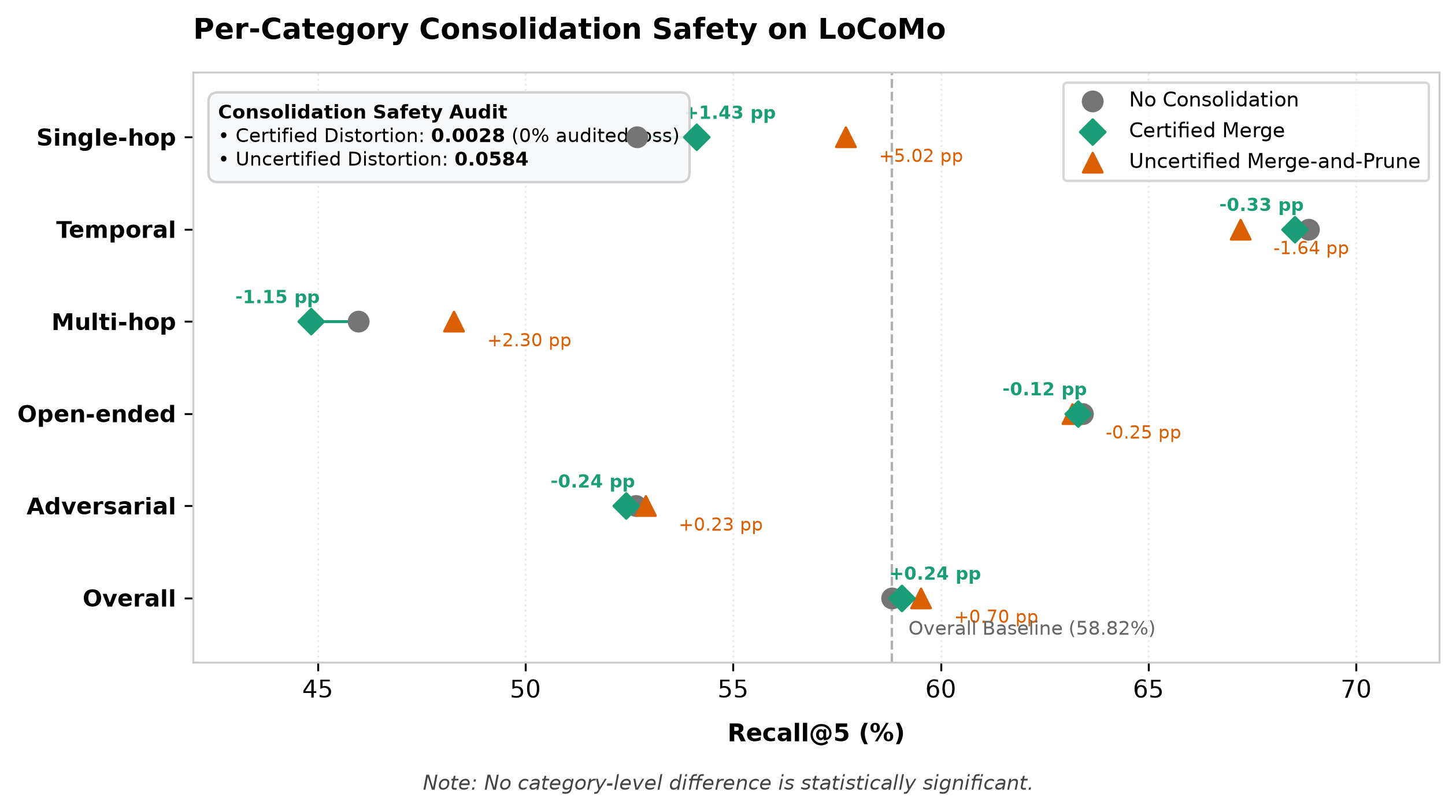}
    \caption{Per-category Rec@5 diagnostic at the retained 20-operation point over 1,905 deduplicated LoCoMo questions. Annotations report changes from each category's no-consolidation baseline. Certified merge preserves overall retrieval quality with low distortion ($\hat{\epsilon}=0.0028$), whereas uncertified merge-and-prune obtains larger category changes with substantially higher distortion ($\hat{\epsilon}=0.0584$). No category-level difference is statistically significant.}
\label{fig:consolidation_category}
\end{figure*}

Figure~\ref{fig:consolidation_category} reports per-category recall changes under the earlier 20-operation policies; these category results were not rerun for the 100-operation extension.

Certified merge changes overall Rec@5 from 58.82\% to 59.06\%, while the uncertified variant reaches 59.52\%. The uncertified increase is accompanied by substantially larger sufficiency distortion, so it is not evidence of safer consolidation.

\paragraph{Scope decisions.} Core certification uses Llama-3.1-8B, while the retained compressive-run documentation assigns merge generation to Llama-3.3-70B-Instruct-Turbo; separately labeled GPT-4o-mini and Llama-3.3-70B diagnostics retain their stated backbones. The hard length gate rejects verbose merge candidates and keeps consolidation conservative. The primary submitted extension uses 100 certified operations, $-1.42\%$ active tokens, 59.08\% FB-TG Rec@5, and 0/100 audited losses. Dominated and budget-mode pruning were independently judged lossy in 83.9\% ($n=174$) and 63.6\% ($n=404$) of operations, respectively, and are excluded from the core. The following 20-operation comparisons are retained only as earlier diagnostics, including the aggressive uncertified merge-and-prune run ($-7.15\%$ tokens, $\hat{\epsilon}=0.0584$).

The complete earlier 20-operation policy rows now appear in the explicitly separated legacy panel of Table~\ref{tab:consolidation_full}; no standalone table duplicates those values.

\subsection{Credit-Ledger Diagnostic}

Table~\ref{tab:credit_ledger} reports attribution credit-ledger stability over 28 matched sleep events at the 100-operation scale. The paired difference remains directional rather than statistically significant.

\begin{table*}[htbp]
\centering
\scriptsize
\setlength{\tabcolsep}{4.5pt}
\begin{tabular}{lcccccc}
\toprule
Variant & $r_{\mathrm{pre}}$ & $r_{\mathrm{post}}$ & $\Delta r$ & 95\% CI on $\Delta r$ & Paired $p$ & Matched sleeps \\
\midrule
Certified merge (primary) & 0.3887 & 0.3831 & $-0.0056$ & $[-0.0210,+0.0098]$ & -- & 28 \\
Uncertified merge+prune & 0.3809 & 0.3639 & $-0.0170$ & $[-0.0385,+0.0045]$ & -- & 28 \\
\midrule
\textit{Difference (uncertified $-$ certified)} & -- & -- & \textbf{$-0.0114$} & \textbf{$[-0.0282,+0.0054]$} & \textbf{.1820} & 28 \\
\bottomrule
\end{tabular}
\caption{Attribution credit-ledger stability over 28 matched sleep events evaluated at the 100-operation scale. $r_{\mathrm{pre}}$ and $r_{\mathrm{post}}$ measure Pearson correlation between item importance and gold-evidence utility before and after sleep. The paired difference is directional ($p=.1820$) and serves as a secondary event-level diagnostic.}
\label{tab:credit_ledger}
\end{table*}

\subsection{Bridge Ablations}

Table~\ref{tab:bridge_ablations} reports the final ten-store bridge ablations at the 100-operation scale. Credit protection and co-citation are neutral safety filters at this operating point, while maximum credit transfer yields the only measurable recall change.

\begin{table*}[htbp]
\centering
\scriptsize
\setlength{\tabcolsep}{4pt}
\begin{tabular}{lcccccccc}
\toprule
Bridge mechanism & $R_A$ (on) & $R_B$ (off) & $\Delta R$ (pp) & 95\% CI & Paired $p$ & $\hat{\epsilon}_A\downarrow$ & $\hat{\epsilon}_B\downarrow$ & $\Delta\hat{\epsilon}$ \\
\midrule
Credit protection & 0.5908 & 0.5908 & $+0.00$ & $[-0.15,+0.15]$ & 1.0000 & 0.0028 & 0.0028 & 0.0000 \\
Maximum transfer & 0.5908 & 0.5888 & \textbf{$+0.20$} & \textbf{$[+0.03,+0.37]$} & \textbf{.0210} & 0.0028 & 0.0034 & \textbf{$-0.0006$} \\
Co-citation proposals & 0.5908 & 0.5908 & $+0.00$ & $[-0.12,+0.12]$ & 1.0000 & 0.0028 & 0.0028 & 0.0000 \\
\bottomrule
\end{tabular}
\caption{Lifecycle bridge ablations at the 100-operation scale ($n=1{,}905$). $R$ is evidence Rec@5; $A/B$ denote enabled/disabled for protection and co-citation, and maximum/zero for transfer. Credit protection and co-citation act as neutral safety filters, whereas maximum credit transfer yields a modest recall gain and lower distortion.}
\label{tab:bridge_ablations}
\end{table*}

\subsection{Pilot Sensitivity Sweeps}
Table~\ref{tab:consolidation_sweeps} reports coarse threshold and scheduling sensitivity on a two-store pilot.

\begin{center}
\centering
\footnotesize
\begin{tabular}{lrrrr}
\toprule
Setting & Rec@5 & $\hat{\epsilon}$ & Tokens & Merges \\
\midrule
$\tau_c=0.4$ & 0.6236 & 0.00198 & 77,105 & 4 \\
$\tau_c=0.6$ & 0.6236 & 0.00198 & 77,105 & 4 \\
$\tau_c=0.8$ & 0.6194 & 0 & 77,204 & 0 \\
$\tau_c=1.0$ & 0.6194 & 0 & 77,204 & 0 \\
Sleep $=4$ & 0.6262 & 0.00178 & 77,105 & 4 \\
Sleep $=8$ & 0.6236 & 0.00198 & 77,105 & 4 \\
Sleep $=16$ & 0.6194 & 0.00260 & 77,158 & 2 \\
\bottomrule
\end{tabular}
\captionof{table}{Two-store sensitivity pilots. These runs document coarse threshold and scheduling behavior and are not treated as final-scale evidence.}
\label{tab:consolidation_sweeps}
\end{center}

\subsection{Aggressive Summarization and DIGEST Diagnostics}

Table~\ref{tab:summarize_all_diagnostic} reports the unconditional whole-store summarization diagnostic and DIGEST end-to-end accuracy.

\begin{center}
\centering
\scriptsize
\textbf{(a) Unconditional whole-store summarization.}\\[2pt]
\begin{tabular}{lrr}
\toprule
Metric & No sleep & Summarize all \\
\midrule
Rec@5 & 0.5882 & 0.7460 \\
Full@5 & 0.3921 & 0.5014 \\
Context recall & 0.7176 & 0.7735 \\
Top-5 tokens & 143.0 & 201.7 \\
Store tokens & 163,680 & 31,688 ($-80.6\%$) \\
$\hat{\epsilon}$ & 0 & --$^{\dagger}$ \\
Probes & 0 & 2,663 \\
\bottomrule
\end{tabular}
\\[6pt]
\textbf{(b) DIGEST end-to-end accuracy on one LoCoMo sample ($n=238$).}\\[2pt]
\begin{tabular}{llcc}
\toprule
System & Backbone & Accuracy & Rec@5 \\
\midrule
Mem0 (same model) & 8B & 35.75\% & -- \\
A-MEM (same model) & 8B & 38.27\% & -- \\
MEMOIR+DIGEST & 8B & 43.28\% & 38.90\% \\
MEMOIR+DIGEST & 70B & 40.76\% & 39.46\% \\
LiCoMemory (live) & 70B & 54.3--54.8\% & -- \\
\bottomrule
\end{tabular}
\captionof{table}{Supplementary synthesis diagnostics. Panel (a) is not treated as a performance claim because evidence identifiers are unioned onto summaries while retrieved context grows by 41\%. $^{\dagger}$Sufficiency distortion is not comparable for summarize-all under identifier-union scoring; the raw estimator returned zero but violates the metric's assumptions. Panel (b) is a single-sample diagnostic and does not support significance claims.}
\label{tab:summarize_all_diagnostic}
\end{center}

The unconditional summarize-all baseline reduced total store tokens from 163,680 to 31,688, but increased the average retrieved top-five context from 143.0 to 201.7 tokens. Its apparent Rec@5 increase from 0.5882 to 0.7460 should not be read as a reliable consolidation win: the operation unions many constituent evidence identifiers onto a few surviving summaries, allowing identifier-intersection metrics to credit a retrieved item even when its text no longer contains a compact, extractable answer. This baseline therefore requires end-to-end QA evaluation or a stricter text-grounded evidence metric before it can support a comparative claim.

\section{Limitations}

TRACE-UCB requires repeated queries over a shared evolving store and therefore cannot improve the first query to a fresh store. Confidence-gated self-feedback depends on calibrated confidence and can reinforce irrelevant evidence when an answer is confidently wrong. Lexical TRACE can miss numeric, short, and paraphrastic evidence, whereas TRACE-CF incurs additional leave-one-out scoring latency. The factorial table uses oracle reference feedback and does not establish the same joint result for the deployed confidence-gated system.

Certified merge is an empirical safety gate rather than a formal guarantee. The primary extension observes 0 losses among 100 audited merges, but its one-sided 95\% operation-level upper bound remains 2.95\%, and operations produced within the same store may be statistically dependent. The 1.42\% reduction is an active-store saving rather than a total-storage saving because originals are archived for rollback. Table~\ref{tab:consolidation_full} reports the recovered $|\mathcal{P}|=1{,}120$ distortion denominator and policy-specific lost-probe counts. The preliminary SUPERSEDE audit observes 0/24 invalid retirements but retains an 11.73\% one-sided upper bound. The credit-ledger comparison is event-level, and its interval crosses zero; it is therefore a secondary diagnostic rather than a system-level safety guarantee.

Results are concentrated on ten LoCoMo stores. LongMemEval-S uses one question per fresh store and therefore evaluates cold-start retrieval and consolidation preservation rather than longitudinal feedback. The submitted 100-operation values are now synchronized across the main paper and supplement, but the frozen per-query FB-TG outputs, accepted-operation log, cross-model audit manifest, exact environment versions, and final repository revision are not present in the supplied workspace. The result therefore remains incompletely reproducible until those artifacts are archived.

\section{Ethics Statement}

Long-term agent memory can retain sensitive personal information across sessions. Deployments should use encryption, access control, retention limits, and data-minimization policies. MEMOIR adds a specific fairness risk: if biased, harmful, or privacy-sensitive memories repeatedly appear in successful trajectories, retrospective importance revision may increase their future retrieval probability. This risk is distinct from generic memory retention because it arises from the feedback loop itself. Practical systems should combine MEMOIR with content filters, user-controlled deletion, and auditing of high-importance memories.

\end{document}